\documentclass[twocolumn,prb,longbibliography,aps,notitlepage,showpacs,amsmath,amssymb,superscriptaddress]{revtex4-1}

\usepackage{float}
\usepackage{times}
\usepackage{bm}
\usepackage{textcomp}
\usepackage{graphicx}
\usepackage{bm}
\allowdisplaybreaks 
\usepackage{braket}
\usepackage{array}
\usepackage[breaklinks]{hyperref}
\usepackage{subdepth}
\usepackage{mathtools}
\usepackage{physics}
\hypersetup{colorlinks=true, linkcolor=blue, citecolor=blue, filecolor=blue, urlcolor=blue}
\newcommand{\RN}[1]{\textup{\uppercase\expandafter{\romannumeral#1}}}
\newcommand{\vbld}[1]{\vectorbold{#1}}
\newcommand{\pd}{\phantom{\dag}}

\begin{document}

\title{Effect of disorder on density of states and conductivity in higher order Van Hove singularities in two dimensional bands}

\author{Anirudh Chandrasekaran}
\affiliation{Department of Physics, Loughborough University, Loughborough LE11 3TU, UK.}

\author{Joseph J. Betouras}
\affiliation{Department of Physics, Loughborough University, Loughborough LE11 3TU, UK.}

\date{\rm\today}

\begin{abstract}
We study systems with energy bands in two dimensions, hosting higher order Van Hove singularities (HOVHS) in the presence of disorder, using standard diagrammatic techniques for impurity averaging. In the clean limit, such singularities cause power-law divergence in the density of states (DOS), and this is expected to strongly affect electronic correlation. In order to analyse the signatures of these singularities in disordered systems, we employ various Born approximations, culminating in the self-consistent (non) Born approximation. Although the divergence of the DOS is smeared, we find that the shape of the DOS, as characterized by the power law tail and the universal ratio of prefactors, is retained slightly away from the singularity. This could help us to understand current and future experiments on materials that can be tuned to host HOVHS. The impurity induced smearing is calculated and analysed for several test cases of singularities. We also study the effects of impurities on electrical conductivity and determine the regimes where the quantitative features of the power law DOS manifest in the conductivity.

\end{abstract}

\maketitle


\section{Introduction}\label{sec:intro}
Novel phases of matter driven by non-trivial topology and geometry of electronic band structure, have been the subject of much interest in recent times. While the various effects of topology have been well studied and documented, new avenues continue to emerge in the investigation of band structure geometries. Pioneering early work by Lifshitz and Van Hove \cite{Lifshitz_1960,VanHove} laid the foundation for a rich path of subsequent explorations uncovering the exotic effects of Fermi surface geometry and its consequences, especially in the context of Fermi surface topological transitions. In Fermi surface topological transitions, the Fermi surface geometry undergoes a sudden change when some parameters in the system are changed. Lifshitz initially studied two particular forms of these topological changes (pocket appearing/disappearing or neck formation/collapse).~\cite{Lifshitz_1960, Abrikosov_1988} Quite often, such transitions happen when the Fermi surface hosts critical points of the dispersion.  

When the gradient of the energy dispersion $\varepsilon_n^{\pd} (\vectorbold{k})$ of the $n^{\mathrm{th}}$ band at some point $\vbld{k}_0^{\pd}$, represented by the Jacobian $\nabla \varepsilon_n^{\pd} (\vectorbold{k_0^{\pd}})$ vanishes, we have a critical point of the dispersion, that could be a maximum, minimum or a saddle. It is often adequate to describe the dispersion around these points with a Taylor expansion to quadratic order, taking a canonical form $\pm k_x^2 \pm k_y^2$. Such extrema are accompanied by a logarithmic divergence in the DOS~\cite{VanHove} at the corresponding energy. There are however, a class of critical points around which the dispersion needs to be Taylor expanded beyond quadratic order. These are points where the determinant of the Hessian of dispersion relation also vanishes, and they are known as higher order critical points, and the corressponding dispersion relation is said to have a higher order Van Hove singularity (HOVHS). An example of HOVHS that is frequently reported in the literature is the cusp singularity having the canonical dispersion $k_x^4 - k_y^{2}$.~\cite{Chandrasekaran, Efremov_2019, Shtyk_2017} 

At a higher order critical point, the Fermi surface becomes singular as happens in a regular Van Hove singularity. HOVHS are also accompanied by power law diverging DOS, often with asymmetric ratio of prefactors above and below the singular energy.~\cite{Chandrasekaran, LiangFu} Signatures of this can be observed experimentally in the tunnelling conductivity.~\cite{magic_VanHove,magic_VanHove2} This property is expected to affect other measurable quantities such as electrical conductivity and charge susceptibility as well. Furthermore, electronic correlation is expected to be enhanced by the large DOS in the vicinity of the HOVHS, potentially leading to novel phases driven by electronic interaction. The kinetic energy of the fermions in the vicinity of those points is comparable or less to the interaction potential energy.

Lifshitz  transitions and associated logarithmically-divergent  Van Hove singularities have been reported in in a variety of materials including
cuprates, iron based superconductors, cobaltates, $\text{Sr}_{2} \text{RuO}_{4}$ and
heavy fermions.~\citep{Aoki,Barber,Bernhabib,Khan,Coldea,Okamoto,Sherkunov-Chubukov-Betouras,Slizovskiy-Chubukov-Betouras,Stewart,Yelland}
There is an even more recent surge of interest in higher order Van
Hove singularities.~\cite{HOVHS1,HOVHS2,HOVHS3,HOVHS4,HOVHS5} Some of the materials where they have been discovered include $\text{Sr}_{3} \text{Ru}_{2} \text{O}_{7}$
where a higher order ($X_9$ with $n=4$) Van Hove singularity was shown to exist in the presence
of an external magnetic field,~\citep{Efremov_2019, Chandrasekaran} while a different higher
order Van Hove saddle has been reported in highly overdoped graphene~\citep{Rosenzweig_2020} and may be quite relevant for the recently observed phases of Bernal bilayer graphene.~\cite{Zhou_2021}

Given the range of exotic physical phenomena that HOVHS promise, it is not a surprise that that HOVHS typically require delicate tuning of parameters in the system to obtain (effected through strain, pressure, twist, bias voltage, etc). This has lately been made possible with the advances in experimental techniques. Although infinitely many such distinct singularities exist, catastrophe theory~\cite{castrigiano,poston,bruce1992curves} guarantees that we are typically likely to obtain only a finite subset of singularities in real systems. For a thorough classification of such HOVHS in two dimensional systems refer~\onlinecite{Chandrasekaran, LiangFu}. Mathematically, these singularities are unstable to certain classes of perturbations that cause the higher order critical points to break into a number of ordinary critical points. Thus, where it is theoretically possible to obtain a HOVHS, it becomes pertinent to investigate the effect of such factors as impurities, that may impede their experimental realization. 

A random distribution of impurities in a crystalline system with a concentration $n_{\text{imp}}^{\pd}$, however low, clearly distorts the lattice periodicity, rendering a simplistic application of the Bloch-band idea questionable. Nevertheless, we expect that for low concentrations, a perturbative, diagrammatic treatment over the clean band theoretic system would be sufficient. There a few such diagrammatic techniques available to treat the problem of randomly distributed impurities, such as quenched averaging, the replica trick, supersymmetry and the Keyldish technique.~\cite{Keldysh, Kamenev, Efetov, replica} In trying to understand the impact of disorder, one typically tries to compute how certain quantities like the spectral function $A_{\vbld{k}}^{\pd} (\omega)$ and the electrical conductivity $\sigma (\omega)$ are affected due to perturbative corrections. The spectral function gives information about the density of states which is a measurable quantity (alongside response functions like various conductivities). For the clean system, these quantities show singular behaviour due to the HOVHS. It is therefore important that we estimate both the qualitative and quantitative changes in the presence of impurities. 

Intuitively, we would expect the signatures of the HOVHS to survive to some extent when the impurity scattering is `weak'. One way to characterize weak scattering is a low concentration of impurities, as compared to concentration of electrons or atoms.~\citep{bruus2004many} More precisely, consider a system with area (or volume) $\mathcal{A}$ having $N_{\text{el}}^{\pd}$ electrons and $N_{\text{imp}}^{\pd}$ impurities so that the concentration of impurities is $n_{\text{imp}}^{\pd} = N_{\text{imp}}^{\pd} / \mathcal{A}$ and concentration of electrons is $n_{\text{el}}^{\pd} = N_{\text{el}}^{\pd} / \mathcal{A}$. We then need $ n_{\text{imp}}^{\pd} \ll n_{\text{el}}^{\pd}$. 

Another important quantity needed to characterise the weakness of impurity scattering is the strength of a single impurity potential denoted by $V(\vbld{x})$. Here we assume that it is short ranged and denote its average value in a unit cell centred around the impurity as $\overline{V}$. Since the zeroth Fourier transform $u_{\vbld{q} = 0}^{\pd}$ is the integral of this short ranged potential over all space, we can express the unit cell average of the potential in terms of it as $\overline{V} \approx u_0^{\pd} / A_{\text{u.c}}^{\pd}$, where $A_{\text{u.c}}^{\pd}$ is the area of the real lattice unit cell. If $E_s^{\,}$ is some energy scale corresponding to the singular dispersion (see Appendix~\ref{app:choosing_E} for a strategy to choose $E_s^{\,}$), then we would expect the impact of the impurity scattering on the singularity to be weak when $\overline{V} \ll E_s^{\,}$, or equivalently
\begin{equation}
u_0^{\pd} \ll A_{\text{u.c}}^{\pd} \, E_s^{\,}.
\end{equation}
This estimate is independent of the condition of validity for the diagrammatic full Born approximation to be derived below. 
It is used to carefully choose the numbers for the numerical calculations that will follow.

In this work, we use the diagrammatic quenched averaging technique to determine the smearing of the power law diverging DOS. To this end, we shall calculate the self energy, and through it the scattering lifetime of electrons in a disordered system hosting a HOVHS in the clean limit. We begin in 
Sec~\ref{sec:systems} by surveying the systems for which the calculations in this paper will be relevant. In Sec~\ref{sec:approxs}, we present in detail, the hierarchy of approximations that we will use to calculate the impurity averaged self energy in a pedagogical manner. We are careful to apply the continuum limit only at the end of calculations, in order to avoid mathematical complications such as analytic continuation of integrals as opposed to finite sums. This is important due to the delicate nature of the problem being studied. After analysing the smearing of the DOS within the various approximations, we proceed to briefly discuss the consequences for electrical conductivity due to HOVHS and impurities in Sec~\ref{sec:conductivity}. We then discuss the relevance to experiments on real materials in Sec~\ref{sec:real_materials}, and summarize and conclude in Sec~\ref{sec:conclusion}. 


%
\section{Systems considered}\label{sec:systems}
The systems of interest in this work are  two-dimensional and quasi two-dimensional Fermi liquid materials (layered) that can be tuned to host a higher order Van Hove singularity near the Fermi level. This can, in practice be achieved in suitable systems by the tuning of a number of parameters, such as the application of pressure, bias voltage, strain, doping and twist angle, to name a few. When a HOVHS occurs near the Fermi level, the low energy dispersion near the Fermi surface can no longer be described adequately by a polynomial of quadratic order (such as an extrema $\xi^{\pd}_{\vb{k}} = \pm (k_x^2 + k_y^2) - \mu$ or a saddle $\xi^{\pd}_{\vb{k}} = \pm (k_x^2 - k_y^2) - \mu$), necessitating a Taylor expansion to higher orders (for example $\xi^{\pd}_{\vb{k}} = \pm (k_x^4 - k_y^2) - \mu$). The primary effect of a HOVHS, within a free electron treatment, is to cause a power law diverging DOS about the energy of the higher order critical point, taking the form
\begin{equation}
g(\epsilon) \sim \begin{cases}
D_+^{\pd} |\epsilon|^{-\nu}, &\epsilon > 0\\
D_-^{\pd} |\epsilon|^{-\nu}, &\epsilon < 0
\end{cases}
\;.
\label{eq:HOVHS_DOS}
\end{equation}
While the actual values of $D_+^{\pd}$ and $D_{-}^{\pd}$ are material specific, their ratio is universal and characteristic of the given singularity class. We document the universal ratio and exponent along with the canonical dispersion for some HOVHS that frequently occur in tight binding models, in Table~\ref{tab:DOS}. The precise definition of the continuum density of states, measured about the singularity (rather than the Fermi level) is
\begin{align}
g(\epsilon) = \int \frac{d^2 q}{(2\pi)^2} \, \delta (\epsilon - \xi_{\vbld{q}}^{\pd} - \mu).
\end{align}\label{eq:DOS_definition}

Throughout the rest of the work, we adopt a convention wherein the zero of low energy dispersion $\xi^{\pd}_{\vb{k}}$ coincides with Fermi level and $\xi^{\pd}_{\vb{k}}$ takes a value equal to $-\mu$ at the HOVHS. However, the DOS $g(\epsilon)$ is defined with respect to the energy \emph{at the} HOVHS, the DOS at the Fermi level given by $g(\mu) \sim |\mu|^{-\nu}$. The reason to define DOS this way is that if it is instead defined about the Fermi level as we do for $\xi^{\pd}_{\vb{k}}$, we will have a factor $g(0)$ rather than $g(\mu)$ to denote the DOS at the Fermi level, and this is obviously an inconvenient choice when analyzing quantities that depend on DOS at the Fermi level, as a function of $\mu$, which measures the closeness of the Fermi level to the singularity. In the limit of $\mu \rightarrow 0$, the Fermi surface develops a point singularity, with an associated power law divergence in the DOS at the Fermi level. 

\begin{table*}
\begin{center}
\begin{tabular}{|c|c|c|}
\hline
Singularity & Dispersion ($\alpha, \beta > 0$) & $g(\epsilon)$ \\
\hline
Fold & $\alpha \, k_x^3 - \beta k_y^2$ &  $\alpha^{-1/3} \, \beta^{-1/2} \,\frac{ 2 \,\Gamma \left( 1/3 \right)}{2 \, \pi^{3/2} \Gamma \left( 5/6 \right)} \left( \Theta(-\epsilon) + \frac{\Theta(\epsilon)}{\sqrt{3}} \right) \, |\epsilon|^{-1/6} $  \\ & & \\
Cusp & $\alpha \, k_x^4 - \beta k_y^2$ & $\alpha^{-1/4} \, \beta^{-1/2} \,\frac{ 2 \, \Gamma^2\left( 5/4 \right)}{\pi^{5/2}} \left( \Theta(-\epsilon) + \frac{\Theta(\epsilon)}{\sqrt{2}} \right) \, |\epsilon|^{-1/4} $  \\ & & \\
\begin{tabular}{@{}c@{}}
Monkey\\
saddle
\end{tabular} & $\alpha ( k_x^3 - 3 \, k_x^{\pd} \, k_y^2 )$ & $\alpha^{-2/3} \, \frac{\Gamma \left( 7/6 \right)}{2 \, \pi^{3/2} \, \Gamma \left( 2/3 \right)} |\epsilon|^{-1/3}$ \\ & & \\
$X_9^{\pd}$ & $\alpha ( k_x^4 +  k_y^4 - 6 \, k_x^2 \, k_y^2)$ & $\alpha^{-1/2} \, \frac{\Gamma \left( 1/4 \right)}{16 \, \pi^{3/2} \, \Gamma \left( 3/4 \right)} |\epsilon|^{-1/2}$ \\
\hline
\end{tabular}
\end{center}
\caption{The density of states per unit area for some of the commonly occurring higher order singularities are tabulated above. We have used a generic form of the ideal dispersions, with arbitrary coefficients $\alpha$ and $\beta$ (We have treated the case where both $\alpha$ and $\beta$ are positive for the sake of simplicity. The discussion can be extended easily to other situations). The presence of non-trivial coefficients simply modifies the power law diverging DOS by an overall constant. The symbol $\Gamma$ specified above is the familiar gamma function that extends the notion of factorials. Keeping in mind the realistic situations where the monkey saddle and $X_9^{\,}$ occur respectively at 3-fold and 4-fold rotationally symmetric points, we have used expressions that respect these symmetries.}
\label{tab:DOS}
\end{table*}

\section{Various approximations}\label{sec:approxs}
\begin{figure*}[t]
\includegraphics[width=1\textwidth]{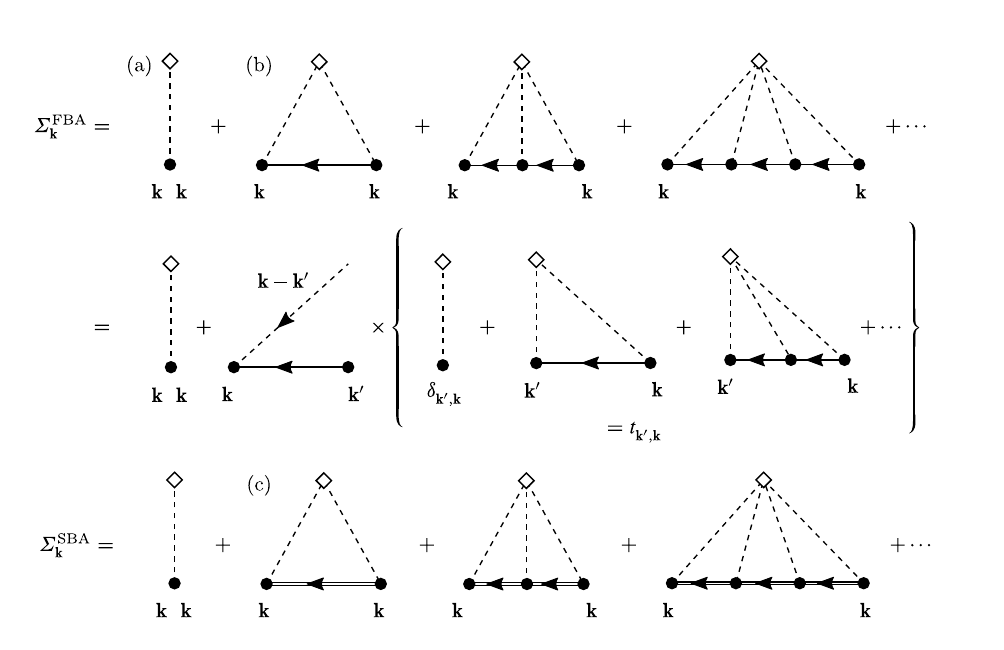}
\caption{Diagrammatic representation of the approximations to the renormalized one particle Green's functions. The simplest is the tree level approximation shown in (a). The next, to one loop, is the first Born approximation (1BA) in (b). The full Born approximation consists of all $O(n_{\text{imp}})$ diagrams. Regarding the self-consistent Born approximation, there are two distinct approximations that are referred to by this name. The simpler one involves diagram (c) while the inclusion of all the succeeding diagrams (which we refer to as the self-consistent Born approximation in this paper) is sometimes referred to as self-consistent non-Born approximation.}
\label{fig:diagrams_Born}
\end{figure*}

Given that the self energy cannot be calculated to arbitrary orders, we employ a series of tractable approximations that retain only particular diagrams or some entire classes of diagrams. The various approximations used below are depicted in Fig~\ref{fig:diagrams_Born}. We briefly note the Feynman rules: dashed impurity lines carrying momentum $\vbld{q}$ contribute an amplitude $u_{\vbld{q}}^{\pd}$ (the Fourier coefficient of a single impurity potential), the solid Fermion lines carrying $(iq_n^{\pd}, \vbld{q})$ are accompanied by the corresponding bare Fermionic propagator $\mathcal{G}_{\vbld{q}}^{0} (iq_n^{\pd})$, the unfilled diamond shaped impurity vertices each contribute a factor of $n_{\text{imp}}^{\pd}$, and momentum is conserved at all vertices. Only internal Fermion momenta are summed over. Note that since impurities are elastic scatterers, no frequency flows through the impurity lines. The precise origin and motivation for these rules can be found in Ref~\onlinecite{bruus2004many}.

Before we proceed further, we explain some aspects of the notation used in this work. We use $\varepsilon_n^{\pd}(\vb{k})$ or simply $\varepsilon (\vb{k})$ to denote the full and actual band dispersion without reference to the Fermi level. In contrast, $\xi^{\pd}_{\vb{k}} = \varepsilon_n^{\pd}(\vb{k}) - \mu$ is the full or series expanded dispersion, adjusted by the chemical potential $\mu$, i.e. at zero temperature, states with negative $\xi^{\pd}_{\vb{k}}$ are occupied while states with positive $\xi^{\pd}_{\vb{k}}$ are unoccupied. For $\mu = 0$, the Fermi level lies exactly at the higher order singularity. The Fermionic Matsubara frequencies are denoted by $k_n^{\pd} =  (2n + 1) \pi / \beta$.

\subsection{Tree level}

At tree level, we only have a single impurity vertex and a single line as in Fig~\ref{fig:diagrams_Born} (a). The tree level self energy is then simply
\begin{equation}
\Sigma_{\vbld{k}}^{\pd} (ik_n^{\pd}) = n_{\text{imp}}^{\pd} \, u_0^{\pd}.
\end{equation}
This contribution is purely real and $\vbld{k}$-independent and provides only a constant shift to the dispersion.

\subsection{First Born approximation}
This is the first loop correction to the self energy, shown in Fig~\ref{fig:diagrams_Born} (b). It takes the value
\begin{equation}
\Sigma_{\vbld{k}}^{\text{1BA}} (i k_n^{\pd}) = \frac{n_{\text{imp}}^{\pd}}{\mathcal{A}} \sum_{\vbld{q}} \left| u_{\vbld{k} - \vbld{q}}^{\pd} \right|^2 \, \frac{1}{i k_n^{\pd} - \xi_{\vbld{q}}^{\pd}}.
\end{equation}
It is the first non-trivial $O(n_{\text{imp}}^{\pd})$ loop correction. Since the sum over $\vbld{q}$ is finite in a finite subsystem, we analytically continue to obtain the retarded self energy as
\begin{equation}
\Sigma_{\vbld{k}}^{\text{1BA}} (\omega) = \frac{n_{\text{imp}}^{\pd}}{\mathcal{A}} \sum_{\vbld{q}} \left| u_{\vbld{k} - \vbld{q}}^{\pd} \right|^2 \, \frac{1}{\omega - \xi_{\vbld{q}}^{\pd} + i\delta}.
\end{equation}
Now, we assume that $u_{\vbld{k} - \vbld{q}}^{\pd} \approx u_0^{\pd}$, convert the $\vbld{q}$ sum to an energy integral with density of states $g(\epsilon)$ and take imaginary part in order to obtain the scattering lifetime
\begin{subequations}
\begin{align}
\frac{1}{2\tau_{\vbld{k}}^{\pd} (\omega)} &= - \text{Im} \, \Sigma_{\vbld{k}}^{\text{1BA}} (\omega), \label{eq:inv_lifetime_a}\\ 
&= \pi \, n_{\text{imp}} \, u_0^2 \int\limits_{\infty}^{\infty} d\epsilon \, g(\epsilon) \, \delta (\omega - \epsilon + \mu), \nonumber \\
& = \pi \, n_{\text{imp}} \, u_0^2 \, g(\omega + \mu),
\end{align}
\label{eq:inv_lifetime}
\end{subequations}
where we have used the density of states per unit area defined in Eq~\ref{eq:DOS_definition}. Notice that the DOS $g(\epsilon)$ used above and in Eq~\ref{eq:HOVHS_DOS} is defined about the energy at the HOVHS i.e $\xi_{\text{VHS}}^{\pd} = -\mu$, so that when converting the $\vb{q}$ sum to an energy integral with a DOS factor, we have $\xi_{\vbld{q}}^{\pd} \rightarrow \epsilon - \mu$.

The density of states will play an important role in our calculations. It is discussed in greater detail in Appendix~\ref{app:DOS}. When $\omega , \mu \rightarrow 0$, the inverse lifetime diverges due to the singular density of states of higher order singularities. The full Born approximation contains \emph{all} the $O(n_{\text{imp}}^{\pd})$ contributions to the self energy, which, diagrammatically speaking, arise from scattering through a single impurity (and therefore a single $n_{\text{imp}}^{\pd}$ contribution).

\subsection{Interlude: the $t$-matrix}
We will need the $t$-matrix for the full Born approximation as it contains information about the scattering of an electron across a \emph{single} impurity. The $t$-matrix is defined diagrammatically in Fig~\ref{fig:diagrams_Born} (within the curly braces $\{ \, \}$). Momentum conservation between ingoing and outgoing momenta is not demanded. Therefore the $t$-matrix has two momenta indices: $t_{\vbld{k}_1, \vbld{k}_2}^{\pd}(i k_n^{\pd})$. As before, only one frequency dependence is needed since the impurity potentials scatter elastically. It is easy to show that the $t$-matrix satisfies a self consistent equation
\begin{align}
t_{\vbld{k}_1, \vbld{k}_2}^{\pd} (ik_n^{\pd}) = & \, n_{\text{imp}}^{\pd} \, u_0^{\pd} \, \delta_{\vbld{k}_1,\vbld{k}_2}^{\pd} \nonumber \\
& + \frac{1}{\mathcal{A}} \sum_{\vbld{q}} u_{\vbld{k}_1 - \vbld{q}}^{\pd} \, \mathcal{G}^0_{\vbld{q}} (i k_n^{\pd}) \, t_{\vbld{q}, \vbld{k}_2}^{\pd}(i k_n^{\pd}).
\end{align}
If we assume that the impurity amplitudes $u_{\vbld{q}}^{\pd}$ are effectively momentum independent, i.e $u_{\vbld{q}}^{\pd} \approx u_0^{\pd}$, then we immediately see that the right hand side of the above equation does not give any non-trivial dependence on $\vbld{k}_{1}^{\pd}$ since the $\vbld{k}_1^{\pd}$ dependence is expressed through the factor $u_{\vbld{k}_1 - \vbld{q}}^{\pd}$ which is now set to a constant. We can thus assume that the $t$ matrix is also momentum independent, which greatly simplifies calculations. 

Analytically continuing $i k_n^{\pd} \rightarrow \omega + i\delta$ and assuming $u_{\vbld{k} - \vbld{q}}^{\pd} \approx u_0^{\pd}$ so that the $t$-matrix depends only on $\omega$ and $\mu$ (with momentum conservation implied) we obtain:
\begin{equation}
t(\omega, \mu) = \frac{n_{\text{imp}}^{\pd} \, u_0^{\pd}}{1 - u_0^{\pd} \frac{1}{\mathcal{A}} \sum_{\vbld{q}} \mathcal{G}_{\vbld{q}}^0 (\omega)}.
\end{equation}
We have made the $\mu$ dependence of the $t$ matrix explicit above. Once again, the inverse area weighted momentum sums can be converted into integrals over energy levels weighted by the DOS. The integrals take the following form that is relevant for both the full Born (FBA) and self-consistent Born (SCBA) approximations
\begin{equation}
I(\omega, \mu, z_0^{\pd}) = \int_{-\infty}^{\infty} d\epsilon \, g(\epsilon) \, \frac{1}{\omega - \epsilon + \mu + z_0^{\pd}},
\label{eq:integral_common}
\end{equation}
where $\text{Im} \, z_0^{\pd} \neq 0$ and the density of states is given by 
\begin{equation}
g(\epsilon) = \begin{cases}
D_+^{\pd} |\epsilon|^{-\nu}, &\epsilon > 0\\
D_-^{\pd} |\epsilon|^{-\nu}, &\epsilon < 0
\end{cases}
\;.
\end{equation}
The integral is evaluated in Appendix~\ref{app:deriv}. The final result reads
\begin{equation}
I(\omega, \mu, z_0^{\pd}) = \frac{2\pi i (\omega + \mu + z_0^{\pd})^{-\nu}}{1 - e^{-i 2\pi \nu}} \, [ e^{-i \pi \nu} D_-^{\pd} - D_+^{\pd} ].
\end{equation}
In evaluating the above integral, the principal branch is used for the fractional power, i.e $(R e^{i\theta})^{-\nu} = R^{-\nu} e^{-i\nu \theta}$ for $\theta \in (0, 2\pi)$. Now for the $t$ matrix, we have $z_0^{\pd} = + i \delta$. This implies that
\begin{equation}
(x + i\delta)^{-\nu} = \begin{cases}
x^{-\nu} (1 - i\delta) & x > 0 \\
|x|^{-\nu} e^{-i\pi \nu} (1 + i\delta) & x < 0
\end{cases},
\end{equation}
where for us $x = \omega + \mu$. Defining $r = D_+^{\pd} / D_-^{\pd}$, we have
\begin{align}
I(\omega, \mu, i\delta) =& 2\pi \, g(\omega + \mu) \, \, \frac{i(e^{-i\pi \nu} - r)}{1 - e^{-i 2\pi \nu}} \nonumber \\
& \times \begin{cases}
\frac{1}{r} (1 - i \delta) & \omega + \mu > 0 \\
e^{-i\pi \nu} (1 + i \delta) & \omega + \mu < 0
\end{cases}.
\end{align}
The $t$ matrix is then given by
\begin{equation}
t(\omega, \mu) = \frac{n_{\text{imp}}^{\pd} \, u_0^{\pd}}{1 - u_0^{\pd} \, I(\omega, \mu, i\delta)}.
\end{equation}
For concreteness, we include the $t$ matrix values for a few of the singularities that have been identified in lattice models in Table~\ref{tab:sings}.

\begin{table*}
\begin{center}
\begin{tabular}{|c|c|c|c|}
\hline
Singularity
&
$D_+^{\pd} / D_-^{\pd}$
&
$\nu$
&
$t(\omega, \mu)$
\\
\hline
Fold
&
$1/\sqrt{3}$
&
$1/6$
&
$\dfrac{n_{\text{imp}}^{\pd} \, u_0^{\pd}}{1 - \pi u_0^{\pd} \, g(\omega + \mu) \,\, \left( \left( \sqrt{3} \right)^{\text{sgn} (\omega + \mu)} - i \right) }$
\\
Cusp
&
$1/\sqrt{2}$
&
$1/4$
&
$\dfrac{n_{\text{imp}}^{\pd} \, u_0^{\pd}}{1 - \pi u_0^{\pd} \, g(\omega + \mu) \,\, \left( \Theta (\omega + \mu) - i \right) }$
\\
\begin{tabular}{@{}c@{}}
Monkey\\
saddle
\end{tabular}
&
$1$
&
$1/3$
&
$\dfrac{n_{\text{imp}}^{\pd} \, u_0^{\pd}}{1 - \pi u_0^{\pd} \, g(\omega + \mu) \,\, \left( \frac{1}{\sqrt{3}} \, \text{sgn} (\omega + \mu) - i \right) }$
\\
$X_9^{\pd}$
&
$1$
&
$1/2$
&
$\dfrac{n_{\text{imp}}^{\pd} \, u_0^{\pd}}{1 - \pi u_0^{\pd} \, g(\omega + \mu) \,\, (1 - i \, \text{sgn} (\omega + \mu) ) }$
\\
\hline
\end{tabular}
\end{center}
\caption{The values of the $t$-matrix for some of the commonly observed singularities, computed assuming momentum independence of the Fourier transform of the impurity potential (i.e $u_{\vbld{q}}^{\,} \approx u_0^{\,}$). This ensures that the $t$-matrix depends only on frequency. The $t$-matrix values tabulated above give the self energy within the full Born approximation, that in turn yields the lifetime that smears the DOS. For the exact values of $D_+^{\,}$ and $D_-^{\,}$ of these singularities, see Table~\ref{tab:DOS}.}
\label{tab:sings}
\end{table*}

\subsection{Full Born approximation}
The full Born approximation (FBA) is the truncation of the perturbation expansion of $\Sigma_{\vbld{k}}^{\pd} (ik_n^{\pd})$ to order $n_{\text{imp}}$, i.e it includes all the scattering across a single impurity. But this is the information contained in the $t$-matrix as we saw above. More precisely
\begin{eqnarray}
\nonumber
\Sigma_{\vbld{k}}^{\text{FBA}} (i k_n^{\pd}) &=&n_{\text{imp}}^{\pd} \, u_0^{\pd} + \frac{1}{\mathcal{A}} \sum_{\vbld{q}} u_{\vbld{k} - \vbld{q}}^{\pd} \, \mathcal{G}_{\vbld{q}}^0 (i k_n^{\pd}) \, t_{\vbld{q}, \vbld{k}}^{\pd} (i k_n^{\pd}) \\
 &=&  t_{\vbld{k}, \vbld{k}} (i k_n^{\pd}).
\end{eqnarray}
Thus, in the Born approximation, the self energy is simply a diagonal element of the $t$-matrix. By assuming that $u_{\vbld{q}}^{\pd} \approx u_{0}^{\pd}$, the $t$-matrix becomes momentum independent and the FBA self energy coincides with the $t$-matrix:
\begin{subequations}
\begin{align}
\Sigma_{\vbld{k}}^{\text{FBA}} (\omega, \mu) =& \, t(\omega, \mu), \\
=& \, \frac{n_{\text{imp}}^{\pd} \, u_0^{\pd}}{1 - u_0^{\pd} \, I(\omega, \mu, i\delta)}.
\end{align}
\end{subequations}

\subsection{Self-consistent Born approximation}
This is an improvement over Born approximation wherein the full Green's function $\mathcal{G}_{\vbld{q}}^{\phantom 0} (ik_n^{\pd})$ is used in place of the bare one $\mathcal{G}_{\vbld{q}}^{0} (ik_n^{\pd})$, in the diagrams of the Born approximation leading to a self consistent relation
\begin{subequations}
\begin{align}
\Sigma_{\vbld{k}}^{\pd} (i& k_n^{\pd}) \nonumber \\ =& \, n_{\text{imp}}^{\pd} \, u_0^{\pd} + \frac{1}{\mathcal{A}} \sum_{\vbld{q}} u_{\vbld{k} - \vbld{q}}^{\pd} \, \mathcal{G}_{\vbld{q}} (i k_n^{\pd}) \, \tilde{t}_{\vbld{q}, \vbld{k}}^{\pd} (i k_n^{\pd}) ,\\
=&  \, n_{\text{imp}}^{\pd} \, u_0^{\pd} \nonumber \\
&+ \frac{1}{\mathcal{A}} \sum_{\vbld{q}} u_{\vbld{k} - \vbld{q}}^{\pd} \, \frac{1}{i k_n^{\pd} - \xi_{\vbld{q}}^{\pd} - \Sigma_{\vbld{q}}^{\pd} (i k_n^{\pd})} \, \tilde{t}_{\vbld{q}, \vbld{k}}^{\pd} (i k_n^{\pd}).
\end{align}
\end{subequations}
Here, we introduce the modified version of the $t$ matrix with the full $\mathcal{G}_{\vbld{q}}^{\phantom 0} (ik_n^{\pd})$ in the place of $\mathcal{G}_{\vbld{q}}^{ 0} (ik_n^{\pd})$ in its expansion. The self consistent relation for this $\tilde{t}$-matrix is
\begin{subequations}
\begin{align}
\tilde{t}_{\vbld{k}_1, \vbld{k}_2}^{\pd} & (ik_n^{\pd}) \nonumber \\
 =& \, n_{\text{imp}}^{\pd} \, u_0^{\pd} \, \delta_{\vbld{k}_1,\vbld{k}_2}^{\pd} \nonumber \\
& + \frac{1}{\mathcal{A}} \sum_{\vbld{q}} u_{\vbld{k}_1 - \vbld{q}}^{\pd} \, \mathcal{G}^{\phantom 0}_{\vbld{q}} (i k_n^{\pd}) \, \tilde{t}_{\vbld{q}, \vbld{k}_2}^{\pd}(i k_n^{\pd}), \\
=& \, n_{\text{imp}}^{\pd} \, u_0^{\pd} \, \delta_{\vbld{k}_1,\vbld{k}_2}^{\pd} \nonumber \\
& + \frac{1}{\mathcal{A}} \sum_{\vbld{q}} u_{\vbld{k}_1 - \vbld{q}}^{\pd} \, \, \frac{1}{i k_n^{\pd} - \xi_{\vbld{q}}^{\pd} - \Sigma_{\vbld{q}}^{\pd} (i k_n^{\pd})} \, \tilde{t}_{\vbld{q}, \vbld{k}_2}^{\pd}(i k_n^{\pd}).
\end{align}
\end{subequations}
The self energy, again, is simply the diagonal element of the $\tilde{t}$ matrix: $\Sigma_{\vbld{k}}^{\text{SCBA}} (i k_n^{\pd}) = \tilde{t}_{\vbld{k}, \vbld{k}} (i k_n^{\pd})$. We now analytically continue $i k_n^{\pd} \rightarrow \omega + i \delta$ and assume that both the $\tilde{t}$ matrix and the self energy depend only on frequency. We can then write
\begin{subequations}
\begin{align}
\Sigma^{\text{SCBA}} (\omega, \mu) =& \, \tilde{t}(\omega, \mu), \\
 =& \, \frac{n_{\text{imp}}^{\pd} \, u_0^{\pd}}{1 - u_0^{\pd} \frac{1}{\mathcal{A}} \sum_{\vbld{q}} \mathcal{G}_{\vbld{q}}^{\phantom 0} (\omega)}, \\
 =& \, \frac{n_{\text{imp}}^{\pd} \, u_0^{\pd}}{1 - u_0^{\pd} \, I(\omega, \mu, -\Sigma^{\text{SCBA}} (\omega, \mu))}.
\end{align}
\end{subequations}
After substituting for $I$, and writing $\nu = m/n$ for integer $m$ and $n$, we can rearrange the resultant expression to obtain a polynomial equation for $\Sigma^{\text{SCBA}} (\omega, \mu) \equiv \Sigma$:
\begin{multline}
\left( \omega + \mu - \Sigma \right)^m \Sigma^n - \frac{1}{(2\pi i u_0^{\pd})^n} \left( \frac{1 - e^{-i2 \pi \nu}}{e^{-i\pi \nu} D_-^{\pd} - D_+^{\pd}} \right)^n \\ \times \left( \Sigma - n_{\text{imp}}^{\pd} u_0^{\pd} \right)^n = 0.
\end{multline} 
However, it is hard to pick the right root for this equation, even if we were able to solve it in particular instances, therefore we adopt an alternate strategy. From the branch chosen to define the function $f(z) = z^{-\nu}$, we have that $\text{arg} \, f(z) \in (-2\pi\nu, 0)$. We can then define $x = \left( \omega + \mu - \Sigma \right)^{-\nu}$. Clearly we need to choose $x$ such that $\text{arg} \, x \in (-2\pi\nu, 0)$. For simplicity let us treat the case where $m=1$. We can get a polynomial equation for $x$:
\begin{equation}
(\omega + \mu) \, \varphi \, x^{n+1} + (n_{\text{imp}}^{\pd} \, u_0^{\pd} - \omega - \mu) \, x^n - \varphi \, x + 1 = 0,
\label{eq:for_x}
\end{equation}
where we have defined a constant characteristic of the singularity
\begin{equation}
\varphi = 2\pi i  \, (u_0^{\pd} \, D_-^{\pd}) \left( \frac{e^{-i \pi \nu} - r}{1- e^{-i 2 \pi \nu}} \right).
\label{eq:const_phi}
\end{equation}
In general, we will have to solve this equation numerically and pick the root $x_0^{\pd}$ with $\text{arg} \, x_0^{\pd} \in (-2\pi\nu, 0)$.

\begin{figure}[h]
 \centering
 \includegraphics[width=\columnwidth]{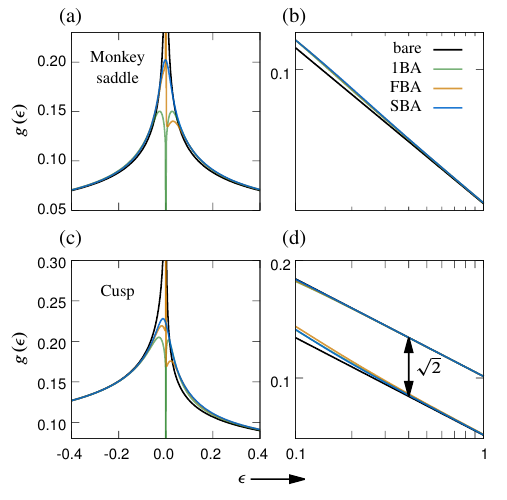}
 \caption{The smeared density of states (DOS) for the monkey saddle and cusp singularities, within the first Born (1BA), full Born (FBA) and self-consistent Born approximations (SCBA) are compared to the DOS of the bare singularity in panels (a) and (c) respectively. We see that the different approximations coincide at large energies $\epsilon$. Near the singularity, the 1BA DOS goes to zero, the FBA DOS diverges, while the SCBA alone captures the finiteness of the smeared singularity. In the panels (b) and (d), we plot these approximations in $\log - \log$ coordinates to make the power law dependence explicit. The asymmetric nature of the cusp singularity manifests as the non-trivial ratio $g(-|\epsilon|) / g(|\epsilon|) = \sqrt{2}$. The rationale behind the choice of numerical parameters used for generating the plot is explained in Appendix~\ref{sec:numerics}.}
 \label{fig:DOS_Borns}
\end{figure}

To compute the smeared DOS, we express it in terms of the spectral function:
\begin{subequations}
\begin{align}
\tilde{g}(\omega) =& \frac{1}{\mathcal{A}} \sum_{\vbld{k}} A_{\vbld{k}}^{\pd} (\omega), \\
=&  \frac{1}{2\pi \mathcal{A}} \sum_{\vbld{k}} \frac{1/\tau_{\vbld{k}}^{\pd}(\omega)}{(\omega - \xi_{\vbld{k}}^{\pd})^2 + 1/4 \tau_{\vbld{k}}^{2}(\omega)}, \\
\approx & \frac{1}{2\pi} \int \frac{d^2 k}{(2\pi)^2} \, \frac{1/\tau_{\vbld{k}}^{\pd}(\omega)}{(\omega - \xi_{\vbld{k}}^{\pd})^2 + 1/4 \tau_{\vbld{k}}^{2}(\omega)}, \\
= & \frac{1}{2\pi} \int\limits_{-\infty}^{\infty} d\epsilon \,\, g(\epsilon) \frac{1/\tau(\omega)}{(\omega - \epsilon)^2 + 1/4 \tau^{2}(\omega)}.
\end{align}
\end{subequations}
Using Eq~\ref{eq:inv_lifetime_a}, where $\tau_{\vbld{k}}^{\pd} (\omega)$ is defined, we can numerically compute the smeared DOS within the various approximations outlined above. We plot the smeared DOS for the cusp and monkey saddle singularities in Fig~\ref{fig:DOS_Borns}.

\subsection{Validity of the expansion}\label{sec:validity}
Before proceeding to analyse the various Born approximations outlined above, we briefly discuss the validity of the diagrammatic technique. We do this in the context of the full Born approximation since the first Born and tree level are contained within this. Furthermore, the topology of the $m^{\text{th}}$ order Born diagram is well known, making it possible to estimate its size. In contrast, the self consistent Born evades a careful analysis of such sorts. 

The $m^{\text{th}}$ order Born diagram contains precisely one impurity vertex $n_{\text{imp}}^{\pd}$, $m$ internal Fermion lines and $m+1$ impurity lines (see the first row of Fig~\ref{fig:diagrams_Born}). The impurity lines simply contribute respective factors $u_{\vbld{q}}^{\pd} \sim u_0^{\pd}$ while the Fermion lines each entail a momentum sum that can be approximated by a momentum integral. Similar to the procedure shown above for the first Born approximation, each such integral roughly yields a factor $g(\mu)$ so that the $m^{\text{th}}$ order Born diagram can be estimated as being proportional to $n_{\text{imp}}^{\pd} \, u_0^{\pd} \, (u_0^{\pd} \, g(\mu))^{m}$. Taking $l$ and $\mathcal{E}$ to denote length and energy units, it is easy to check that this expression has a dimension of energy $\mathcal{E}$, since $[n_{\text{imp}}^{\pd}] = l^{-2}$, $[u_0^{\pd}] = \mathcal{E} \, l^{2}$ and $[g] = \mathcal{E}^{-1} \, l^{-2}$ (since $u_0^{\pd}$, the zeroth Fourier coefficient is simply the integral of the scattering potential over all space and we use the density of states per unit area. See Ref~\onlinecite{bruus2004many} for more details). As the energy factor $n_{\text{imp}}^{\pd} \, u_0^{\pd}$ is common to all diagrams, the dimensionless parameter determining the validity of the expansion is $u_0^{\pd} \, g(\mu)$. Thus, we would expect the full Born approximation to be valid in the regime $u_0^{\pd} \, g(\mu) \ll 1$. Although we are unable to make such precise estimates in the case of self consistent Born approximation, we can compute the hierarchy of energies for each of the above approximations, that is the energy beyond which SCBA coincides with FBA, and the energy beyond which FBA coincides with 1BA. This is done below.

\subsection{Analysis of the self energy in self consistent Born approximation}\label{sec:SBA_analysis}
The self energy in the SCBA satisfies the following relation as seen from the discussion above
\begin{equation}
\Sigma^{\text{SCBA}} (\mu) = \frac{n_{\text{imp}}^{\pd} \, u_0^{\pd}}{1 - \varphi \, (\mu - \Sigma^{\text{SCBA}} (\mu))^{-\nu}},
\end{equation}
with the constant $\varphi$ defined in Eq~\ref{eq:const_phi}. When we are in the regime where $|\mu| \gg |\Sigma^{\text{SCBA}}(\mu)|$, we can essentially ignore $\Sigma^{\text{SCBA}}(\mu)$ to write $(\mu - \Sigma^{\text{SCBA}}(\mu))^{-\nu} \approx \mu^{-\nu}$. Let us focus on $\mu > 0$. It is straightforward to extend the discussion to $\mu < 0$. We now pull out a factor $D_+^{\pd} = r \, D_-^{\pd}$ from $\varphi$ to obtain
\begin{equation}
\Sigma^{\text{SCBA}}(\mu) \approx \frac{n_{\text{imp}}^{\pd} \, u_0^{\pd}}{1 - u_0^{\pd} \, \zeta \, g(\mu)},
\end{equation}
with the constant $\zeta$ being characteristic of the singularity and independent of impurity concentration and strength
\begin{equation}
\zeta = \frac{2\pi i}{r} \left( \frac{e^{-i\pi \nu} - r}{1 - e^{-i2\pi \nu}} \right).
\end{equation}
This approximate expression for self energy coincides with that of the full Born approximation so that the condition for SCBA to pass into FBA is $|\mu| \gg |\Sigma^{\text{SCBA}}(\mu)|$. The inverse lifetime is now obtained by taking the imaginary part
\begin{subequations}
\begin{align}
\tau^{-1}(\mu) =& \, -2 \, \text{Im} \, \Sigma^{\text{SCBA}} (\mu) , \\
=& \frac{2 \, n_{\text{imp}}^{\pd} \, u_0^{2} \, g(\mu) \,\, \text{Im} \, \zeta^*}{\left| 1 - u_0^{\pd} \, \zeta \, g(\mu) \right|^2}.
\label{eq:FBA_lifetime}
\end{align}
\end{subequations}
Now for the standard singularities, we can verify explicitly that $\zeta$ is a small number taking a numerical value near $1$ (for example, $|\zeta| = \sqrt{2} \, \pi$ for the cusp). Therefore, when we have $u_0^{\pd} \, g(\mu) \ll 1$, we can ignore it in the denominator and we have
\begin{subequations}
\begin{align}
\tau^{-1}(\mu) \approx & \, (2 \, n_{\text{imp}}^{\pd} \, u_0^{2} \,\, \text{Im} \, \zeta^*) \, g(\mu), \\
= & \, (2 \, n_{\text{imp}}^{\pd} \, u_0^{2} \, \pi) \, g(\mu).
\label{eq:u0_is_small}
\end{align}
\end{subequations}
To get to the last step we use the fact that $\text{Im} \, \zeta^* = \pi$ for any possible value of $\nu$ and $r$. This is the inverse lifetime given by the first Born approximation as well, as can be checked by comparing to Eq~\ref{eq:inv_lifetime}. Therefore, the condition $u_0^{\pd} \, g(\mu) \ll 1$ allows us to transition from FBA to 1BA. Notice that this condition was also shown to validate the FBA in Sec~\ref{sec:validity}. To summarize, we have
\begin{equation}
\text{SCBA} \xrightarrow{|\mu| \gg |\Sigma^{\text{SCBA}}(\mu)|} \text{FBA} \xrightarrow{u_0^{\pd} \, g(\mu) \ll 1} \text{1BA}.
\end{equation}
Furthermore, in the 1BA regime, $\tau^{-1} (\mu) \propto g(\mu)$. 

Let us examine the case $u_0^{\pd} \, g(\mu) \gg 1$ but with $|\mu| \gg |\Sigma^{\text{SCBA}}(\mu)|$ in some detail. While applicability of the FBA may be questionable here, we rely on the SCBA for analysis. Since $u_0^{\pd} \, g(\mu) \gg 1$, we ignore the $1$ in the denominator of Eq~\ref{eq:FBA_lifetime} and obtain 
\begin{equation}
\tau^{-1}(\mu) \approx \frac{2 \, n_{\text{imp}}^{\pd} \,\, \text{Im} \, \zeta^*}{|\zeta|^2 \, g(\mu)} = \frac{2 \, n_{\text{imp}}^{\pd} \, \pi}{|\zeta|^2 \, g(\mu)}.
\label{eq:u0_is_large}
\end{equation}
Thus, in the regime where $|\mu| \gg |\Sigma^{\text{SCBA}} (\mu)|$ and $u_0^{\pd} \, g(\mu) \gg 1$ the lifetime, rather than the inverse lifetime has the signature of the DOS, i.e $\tau (\mu) \propto g(\mu)$. Our numerical studies indicate that the first condition is always satisfied even slightly above $\mu = 0$. In fact the real and imaginary parts of the SCBA self energy approach finite constants in the large $\mu$ limit. Furthermore, the finiteness and smallness of $\text{Im} \, \Sigma^{\text{SCBA}}$ in the respective regimes of interest is consistent with Eq~\ref{eq:u0_is_small} and Eq~\ref{eq:u0_is_large}. 

Therefore, depending on the strength of the scattering potential $u_0^{\pd}$ as compared to the inverse of the density of states $(g(\mu))^{-1}$, either the resistivity (proportional to $\tau^{-1}$) or conductivity (proportional to $\tau$) can show quantitative signatures of the underlying singularity. Notice that this result is \emph{independent} of the concentration of the impurities. While we may object that the condition $u_0^{\pd} \, g(\mu) \gg 1$ renders FBA inapplicable, we have relied here on the SCBA rather than FBA to draw this conclusion. We should be cautious in carrying over the validity estimates of FBA, an essentially perturbative approximation, over to the SCBA, that is a partly non-perturbative treatment. The precise relation between conductivity/resistivity and the lifetime is elaborated in the section below.

\section{Conductivity}\label{sec:conductivity}
The DOS near the Fermi level can be found experimentally by measuring the tunnelling conductance. While this yields direct information about the possible underlying band singularity, we would also like to explore the effects of the singularity on a bulk property of the material, namely the electrical conductivity $\sigma$. As we show in Eq~\ref{eq:conductivity} (see Appendix~\ref{app:finite}), the conductivity, under suitable conditions, becomes proportional to the lifetime of electrons at the Fermi surface. From the preceding discussion, it is clear that the lifetime (or the inverse lifetime) itself becomes proportional to the power-law DOS in certain regimes, particularly when the concentration and strength of the impurities is `low'. The measurement of the DC conductivity as a function of the chemical potential $\mu$, will then serve as a bulk probe of the singularity. 

Here we summarize the interesting features of conductivity in 2D systems hosting a HOVHS near the Fermi surface. We have identified two possible signatures of HOVHS that may occur in the conductivity, namely a direct and inverse dependence of the conductivity on the DOS at the Fermi level. We also list the associated requirements on the concentration and strength of impurities for these signatures to manifest. Firstly, let us recall the expression for the generally asymmetric power law DOS as a function of energy (or Fermi level) for a HOVHS.

\begin{equation}
g(\mu) = (D_{-}^{\pd} \, \Theta(-\mu) + D_{+}^{\pd} \, \Theta(\mu)) \, |\mu|^{-\nu},
\end{equation} 
where $\Theta (x)$ is the step function. The scaling of the conductivity as a function of the Fermi energy near a HOVHS is then
\begin{subequations}
\begin{align}
\sigma(\mu) & \propto g(\mu), \, \text{if} \,\, |\mu| \gg |\Sigma^{\text{SCBA}}(\mu)| \, \text{and} \,\, u_0^{\pd} g(\mu) \gg 1, \\
\sigma(\mu) & \propto (g(\mu))^{-1}, \, \text{if} \,\, |\mu| \gg |\Sigma^{\text{SCBA}}(\mu)| \, \text{and} \,\, u_0^{\pd} g(\mu) \ll 1.
\end{align}
\end{subequations}
As mentioned earlier, in our numerical studies on a few tight-binding models hosting HOVHS, we found that the first condition is always satisfied for  the chemical potential $\mu$ lying close to, but away from the singularity, i.e slightly above or below it. As for the dimensionless parameter $u_0^{\pd} \, g(\mu)$, we found that it was always small, but not large in the weak impurity scattering regime. Recall that the condition for weak impurity scattering was $u_0^{\pd} \ll A_{\text{u.c}}^{\pd} \, E_s^{\,}$, where $A_{\text{u.c}}^{\pd}$ is area of the real lattice unit cell and $E_s^{\,}$ is some suitable energy scale/window corresponding to the singularity, such as a band width. Therefore, in the models explored numerically in this work, only the second scenario was achieved (i.e the inverse dependence of conductivity on the DOS), but not the first (where there would be direct proportionality between conductivity and DOS). However we do not preclude the possibility of the first scenario from occurring in other systems hosting HOVHS. Further exploration is needed to determine if this theoretical possibility can ever occur in practice.

The conditions for the conductivity to be either proportional to the DOS (i.e $\sigma (\mu) \propto g(\mu)$) or inversely proportional to it (i.e $\sigma (\mu) \propto (g(\mu))^{-1}$), were derived under somewhat restrictive assumptions, particularly, the `weak' $\vb{k}$ and $\omega$ dependence of the lifetime $\tau$ that arises due to impurity averaging (see Appendix~\ref{app:finite}). Even within the Born schemes outlined above, we expect the $\vb{k}$ and $\omega$ dependence of $\tau$ to lead to non-trivial anisotropy for both the DC and AC conductivities. A calculation to carefully work out such details is beyond the scope of the present work. Nevertheless, we hope that the conclusions presented here adequately capture the essential features of the variation of the conductivity with chemical potential near a band singularity.

\subsection{Strong disorder and interaction}
Two possibilities beyond the `weak' disorder and free electron treatment in this work, demand closer scrutiny: the presence of electronic correlation in addition to disorder and the case of strong impurity scattering. These considerations are experimentally quite relevant and are expected to lead to non-trivial physics. Therefore, the conjunction of a singular underlying band structure with electronic correlation and/or strong impurity scattering merits a careful investigation.

The DFT derived band structure that one attempts to describe using tight-binding models already incorporates electronic interaction at a basic level. For strongly correlated materials where this level of description does not suffice, we may hope to employ diagrammatic techniques to perturbatively calculate quantities of interest. The joint diagrammatic treatment of disorder and perturbation can be achieved using a quenched disorder scheme or more involved schemes such as the Keldysh technique. The former has been successfully employed in the past within the random phase approximation (RPA), to describe the electronic conductivity in Fermi liquids~\citep{bruus2004many,mahan,coleman}. An important yet elementary point of departure from the discussion of conductivity presented here would be the use of an RPA renormalized impurity potential in the Born calculations leading to the conductivity. \\
The various Born approximations used in this work correspond to scattering across a single impurity. At low temperatures and higher concentration of impurities, multi-impurity scattering becomes important along with electronic correlation. This may lead to universal conductance fluctuation and weak localization.

In the case of strong disorder, for a non-interacting or moderately interacting system, what is expected is Anderson localisation \cite{anderson1958}. This physics remains intact at the single-particle level. For the case of strong interactions though, in the presence of strong disorder the possibilities of a glassy behavior versus Mott gap \cite{Schwab_Chakravarty} and many-body localisation need to be investigated further. In addition, the possibility for Griffith's phases \cite{Vojta1, Vojta2} in the strong disorder limit, given the presence of singular density of states, must be also studied further. This is left for future work.

\section{Relevance to real materials}\label{sec:real_materials}
As we noted in the introduction, recently there has been an increase of the number of new materials that are known to exhibit HOVHS. This necessitates the further development of the theoretical machinery describing these singularities. Some of the notable materials which were found to host higher order singularities include twisted bilayer graphene at magic angle~\cite{magic_VanHove,magic_VanHove2}, bilayer transition metal dichalcogenide~\cite{HOVHS3}, Sr$_3$Ru$_2$O$_7$ and more recently, the quasi two-dimensional Kagome superconductors,\cite{kang2022twofold,hu2021rich} with a unique interplay between lattice geometry, topology and flat bands. Earlier too, the existence of HOVHS was inferred in some materials (including high T$_c$ superconductors), where they were referred to as `extended' Van Hove singularities~\citep{evHs1,evHs2,evHs3,evHs4,evHs5,evHs6,evHs7}. Although the precise nature of the dispersion was not well described in those instances, it was recognized that these extended saddles were `flatter' than the regular VHS and were accompanied by power law diverging DOS in 2D. 

Energy dispersion measurements made using angle resolved photoemission spectroscopy (ARPES) have been the primary technique to diagnose the presence of HOVHS. In some cases, the energy bands have even been fit with polynomial low energy theories to demonstrate a dispersion going beyond quadratic order~\cite{kang2022twofold}. Nevertheless, other indirect techniques such as tunnelling conductivity measurements of DOS can also reveal the presence of HOVHS. This is particularly relevant to our results, since we show that the power law tail and asymmetric ratio of prefactors can survive in the presence of disorder. Thus, we might expect to discern signatures of HOVHS in the measurements of the DOS in real materials that are typically characterized by the presence of impurities and finite sample sizes. In the case of TBG, it was indeed a measurement of the DOS by tunnelling conductivity~\cite{magic_VanHove} that revealed an asymmetric power law tail closely resembling the cusp singularity (see Table~\ref{tab:DOS}). Similar measurements on some of the other materials identified or conjectured to host HOVHS (based on ARPES measurements), could provide a conclusive diagnosis, apart from helping us to unambiguously identify the underlying singularity (Since they are uniquely identified by the exponent and ratio of prefactors. See Ref.~\onlinecite{Chandrasekaran}).

Furthermore, studying the dependence of the conductivity and the DOS on doping might help us better understand the correlated electron mechanisms behind the unconventional phases observed in some of these materials. It might also reveal the origin of the HOVHS due to the coalescing of a set of ordinary Van Hove points under tuning, and clarify the role played by the HOVHS in driving the emergence of exotic many-body phases. Such experimental explorations would have to be augmented by subsequent theoretical calculations that expand on the material presented in the current work.

\begin{figure}[h]
 \centering
 \includegraphics[width=\columnwidth]{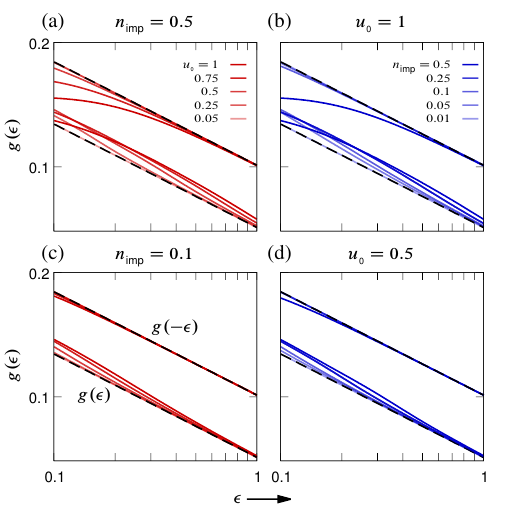}
 \caption{Within the self-consistent Born approximation, the impurity smeared density of states is finite at the singularity, but has a power law tail like the bare singularity. The resemblance to the bare singularity improves as the concentration of impurities $n_{\text{imp}}^{\,}$ and the strength of the scattering potential $u_0^{\,}$ become smaller. This is depicted in the figure above where, in panels (a) and (c), we fix $n_{\text{imp}}^{\,}$ and vary $u_0^{\,}$, while in panels (b) and (d), we fix $u_0^{\,}$ and vary $n_{\text{imp}}^{\,}$. We observe that the smeared DOS and bare DOS become indistinguishable at either low concentrations or weak scattering strength. The choice of numerical parameters used is explained in Appendix~\ref{sec:numerics}. We note that the condition $u_0^{\,} \, g(\mu) \ll 1$, mentioned in Sec~\ref{sec:SBA_analysis}, is trivially satisfied in the situations treated above, with $\epsilon$ instead of $\mu$.}
 \label{fig:DOS_SBA}
\end{figure}

\begin{figure*}[t]
\includegraphics[width=1\textwidth]{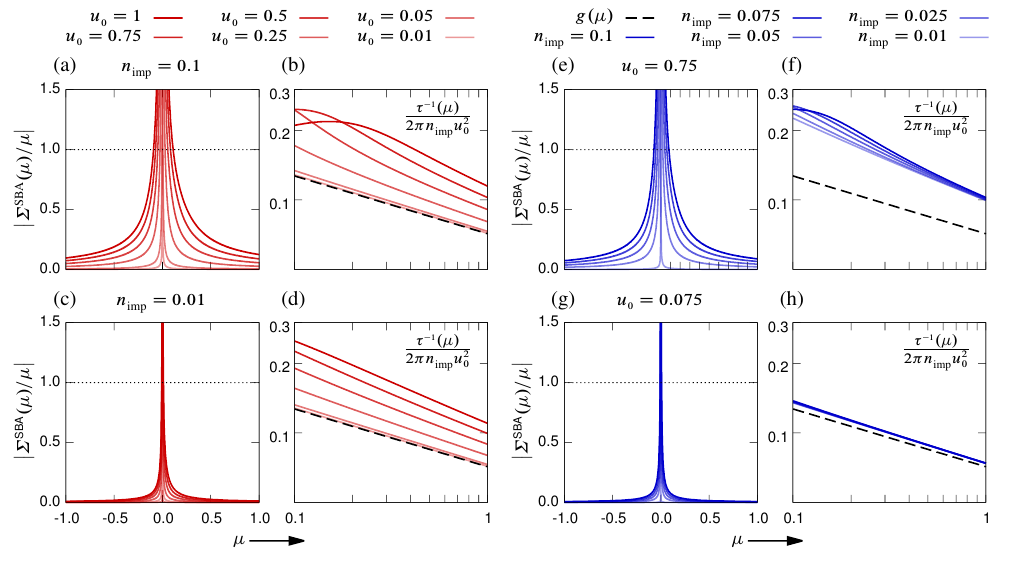}
\caption{Signatures of the power law diverging DOS can manifest in the inverse life-time $\tau^{-1}(\mu)$ of the electrons at the Fermi level. The primary requirement for this, derived in Sec~\ref{sec:SBA_analysis}, is that the self energy within the self-consistent Born approximation has to be much smaller than the chemical potential $\mu$, that measures the energy distance of the Fermi level from the singularity. Mathematically this reads $| \Sigma^{\text{SCBA}}(\mu) / \mu | \ll 1$. In addition to this, when the product $u_0^{\,} \, g(\mu)$, of the `strength' of the scattering potential and the density of states at the Fermi level is much smaller than unity (see Fig~\ref{fig:DOS_SBA}), the inverse lifetime of the electrons at the Fermi level becomes approximately proportional to $g(\mu)$. In the pairs of panels (a), (b) and (c), (d), we fix the impurity concentration and vary $u_0^{\,}$ in a system hosting the monkey saddle, perturbed by impurities. While the condition for power-law signatures to appear in $\tau^{-1}(\mu)$ is analysed graphically in the left panels (a) and (c), the inverse lifetime, weighed by appropriate factors is compared to the DOS in the right panels (b) and (d). A similar scheme with $u_0^{\,}$ fixed and varying $n_{\text{imp}}^{\,}$ is shown in the pairs of panels (e), (f) and (g), (h). We see that the signatures of the power law DOS are seen well in $\tau^{-1}(\mu)$, particularly as either $u_0^{\,}$ or $n_{\text{imp}}^{\,}$ become small. The tight binding model where the singularity is obtained, including the choice of numerical of parameters, is discussed in Appendix~\ref{sec:numerics}.}
\label{fig:sigma_and_tauInv}
\end{figure*}

\section{Discussion}\label{sec:conclusion}
In the previous sections, we have outlined the mathematical procedure involved in analysing higher order singularities in two dimensional bands, in the presence of impurities. We have elucidated various practical approximations (the hierarchy of Born approximations) that make an analytic and numerical analysis tractable. Furthermore, keeping in mind the requirements of numerical calculations and the intuitive analysis of weak impurity scattering made in the preceding sections, we worked out a scheme to choose and/or analyse the impurity concentration and strength in the context of lattice tight binding models (see Sec~\ref{sec:intro} and Appendices~\ref{app:choosing_E},~\ref{sec:numerics}). This broad framework allows us to treat various lattice models hosting HOVHS and perform pertinent calculations. In this section, we analyse the results of such a calculation. 

The primary quantity of interest when dealing with a band singularity is the DOS, which provides a measurable diagnosis of the singularity. To this end, we analyse the impurity induced smearing of DOS due in two HOVHS, the cusp singularity (with dispersion $k_x^4 - k_y^2$) and the monkey saddle singularity (having dispersion $k_y^3 - 3 \, k_y^{\pd} \, k_x^2$). While an analysis of all of the seventeen singularities part of the catastrophe theory classification~\citep{Chandrasekaran} is possible, we restrict the presentation to these two for brevity, since the qualitative features are similar. In fact, the monkey saddle being an odd, three-fold rotationally symmetric singularity with particle-hole symmetry and symmetric DOS and, the cusp being an even, two-fold rotationally symmetric singularity having asymmetric DOS and no particle hole symmetry are reasonable representatives of the rest of the singularities.

From Fig~\ref{fig:DOS_Borns}, we see that the smeared DOS under various Born approximations coincides with the bare DOS at large energies (away from the singularity) while somewhat closer to the singularity, the full Born and self-consistent Born coincide and very close to the singularity, only the SCBA has a finite peak. All this sits well with the analysis of Sec~\ref{sec:SBA_analysis}. As the $\log - \log$ plot reveals, the power-law tail of the DOS and the asymmetric ratio of prefactors are present even in the smeared DOS (of the various approximations) under suitable conditions. This important fact addresses one of our primary questions as to whether any features of the singularity will survive impurity averaging. Furthermore, as Fig~\ref{fig:DOS_SBA} reveals, the smeared DOS in SCBA reflects the bare DOS better and better as the concentration and scattering strength of the impurities decreases.

In Sec~\ref{sec:SBA_analysis}, we laid out the condition under which the inverse lifetime of the electrons at the Fermi level may be proportional to the DOS. This condition along with the inverse lifetime itself is analysed graphically in Fig~\ref{fig:sigma_and_tauInv}. We observe that so long as the impurity strength and concentration remain small, the inverse lifetime does become proportional to the bare DOS slightly away from the singularity.
Since the inverse lifetime plays an important role in the electrical conductivity, we subsequently analysed the conductivity in Sec~\ref{sec:conductivity}. As mentioned above, the integrals multiplying the lifetime in the expression for conductivity (in Eq~\ref{eq:conductivity}) are shown to be finite and non-zero in the $\mu \rightarrow 0$ limit in Appendix~\ref{app:finite}, allowing us to conclude that the signatures of the singularity may potentially appear in the conductivity as well. 

In this paper, we made a diagrammatic analysis of the problem of higher order singularities in unclean two-dimensional systems, using well known approximations. We found that in so far as impurity scattering is weak, it allows much of the quantitative signatures of the HOVHS to survive in measurable physical quantities such as DOS and electrical conductivity. While potential comparison to real materials and experiments may be possible even at this level of description, a logical and necessary next step would be to go beyond the approximations applied in this work. The effect on other response functions like magnetic and thermal susceptibilities is left for future work. An investigation of such a nature would entail the discussion of the combined effects of both disorder and electronic interactions, and this may lead to novel phases driven by instabilities, apart from non-trivial renormalizations of the Fermi surface geometry itself. Lastly, another interesting direction is to extend these calculations to three dimensional systems where there is a diverging DOS (The three-dimensional avatars of the point singularities analysed in this paper do not cause diverging DOS. However line singularities in three dimension can lead to divergences in the DOS~\citep{igoshev2019electron}). For power law diverging DOS in three dimension, we expect the situation to be qualitatively similar to the two dimensional case analyzed in this paper. Nevertheless, this has to be established carefully, and further calculations might be needed to determine the fate of logarithmic divergences in three dimensional DOS under impurity averaging. We hope to explore these problems in a future work.


\acknowledgements{}
We would like to thank Claudio Chamon, Siddhant Das, Dima Efremov, Mark Greenaway, Garry Goldstein and Alex Shtyk for useful discussions.
The work has been supported by the EPSRC grant EP/T034351/1.

\appendix

\section{Choosing the energy scale $E_s^{\,}$}\label{app:choosing_E}

The energy scale $E_s^{\,}$ is chosen to give a rough energy window around the singularity, where the polynomial dispersion of the pristine singularity (and its power law DOS) describe the original dispersion `adequately'. There are a number of reasonable strategies we can employ to choose this. An obvious strategy is to look for other critical points around the higher order critical point. These must necessarily exist since the dispersion is periodic. The minimum of the absolute difference between the energy of the singularity and the energies of other critical points, provides a natural scale at which the polynomial singularity fails to describe the actual dispersion. Mathematically, if the singularity is located at $\vbld{k}_0^{\pd}$,
\begin{equation}
E_s^{\,} = \inf\limits_{\substack{\vbld{k}', \, \text{s. t} \\ \nabla \xi_{\vbld{k}'}^{\,} = 0}} \left| \xi_{\vbld{k}_0}^{\pd} - \xi_{\vbld{k}'}^{\pd} \right|,
\end{equation}
where $\vbld{k}'$ lies in some neighborhood of the singularity that is sufficiently small, so as to avoid critical points that are degenerate to the HOS. A simplification of this procedure is to restrict the search to only the high symmetry points of the BZ (These are already constrained by symmetry to be critical points. See Ref~\onlinecite{Chandrasekaran}). Another strategy is to use the coefficient of the first non-zero higher order correction in the Taylor expansion to compute the energy at which this term becomes important in comparison to the pristine singularity. For example, the monkey saddle expanded to quartic order takes the form $\xi_{\vb{k}}^{\pd} = a \, k^3 \cos (3\varphi) + b \, k^4 - \mu$. Here, $|a^4/b^3|$ has dimensions of energy and provides a natural scale at which the cubic part of the dispersion fails to adequately the full dispersion with the quartic correction.

\section{Discussion on the density of states}\label{app:DOS}
\subsection{Scaling and DOS}\label{app:scaling}
The power law dependence of the DOS in a higher order singularity can be obtained by rescaling the momentum integration variables $k_x^{\pd}$ and $k_y^{\pd}$ in the following integral
\begin{equation}
g(\epsilon) = \int\limits_{\text{u.c}} \frac{d^2 k}{(2\pi)^2} \,\, \delta (\xi_{\vbld{k}} - \epsilon),
\end{equation}
where the integration is over a unit cell centred at the singularity, and the precise scaling transformation $k_x^{\pd} = \epsilon^{p} u$ and $k_y^{\pd} = \epsilon^{q} v$ is characteristic of the respective singularities. Doing so we obtain the power law divergent DOS taking the form $g(\epsilon) \sim \epsilon^{p+q-1}$. 

Such a scaling transformation of course modifies the domain of integration. But the leading order divergent term can be obtained by extending the domain of integration over the entire $k$-plane, and by doing so we pick up only a finite error~\citep{Chandrasekaran}. For the prefactor above and below the singular energy, we still have to evaluate the integral. (See Refs~\onlinecite{Chandrasekaran} and~\onlinecite{LiangFu} for such calculations). For the pristine forms of the singularities (i.e the canonical forms, such as $k_x^{4} - k_y^2$ etc), the precise values of the coefficients can been evaluated. In real systems however, we expect the each of the terms in the dispersion to have non-trivial coefficients, for example $\xi_{\vbld{k}}^{\pd} = \alpha \, k_x^4 - \beta \, k_y^2$. Let us assume for simplicity that $\alpha$ and $\beta$ are positive. Appropriate modifications can be made in the forthcoming procedure when either or both of them are negative. We first do a slightly different scale transformation $k_x^{\pd} = \alpha^{-p} \epsilon^{p} \, u$ and $k_y^{\pd} = \beta^{-q} \epsilon^{q} \, v$ and the density of states for this case, denoted by $\tilde{g} (\epsilon)$, differs from the DOS for the pristine singularity $g(\epsilon)$ by simply an overall constant, i.e
\begin{equation}
\tilde{g} (\epsilon) = \alpha^{-p} \beta^{-q} \, g(\epsilon).
\end{equation}
We state the exact expressions for the DOS of the singularities treated in the paper, near the critical energy in Table~\ref{tab:DOS}.

\subsection{Signature of DOS: $\tau (\mu)$ vs $\tau^{-1}(\mu)$}
In Sec~\ref{sec:SBA_analysis}, we laid out the conditions to be satisfied so that either $\tau(\mu)$ or $\tau^{-1} (\mu)$ may show signatures of the power law diverging DOS. Here we work out the energy window where the signatures may be observed. For $\tau (\mu) \sim g(\mu)$, we need $u_0^{\pd} \, \tilde{g} (\mu) \gg 1$ so that we have
\begin{equation}
\mu_{\text{min}}^{\pd} \ll \mu \ll \text{min} \left\{ \left( \frac{D_+^{\pd} \, u_0^{\pd}}{\alpha^{p} \beta^{q}} \right)^{1/\nu} , E_s^{\,} \right\}.
\end{equation}
For $\tau^{-1} (\mu) \sim g(\mu)$ we instead need 
\begin{equation}
\text{max} \left\{ \mu_{\text{min}}^{\pd} , \left( \frac{D_+^{\pd} \, u_0^{\pd}}{\alpha^{p} \beta^{q}} \right)^{1/\nu} \right\} \ll \mu \ll E_s^{\,},
\end{equation}
where $E_s^{\,}$ is the energy scale described in Appendix~\ref{app:choosing_E}, where the description in terms of the pure singularity breaks down. Lastly, $\mu_{\text{min}}^{\pd}$ is the scale at which 
\begin{equation}
\mu_{\text{min}}^{\pd} \approx \left| \Sigma^{\text{SCBA}} (\mu_{\text{min}}^{\pd}) \right|.
\end{equation}
(Recall that $\mu_{\text{min}}^{\pd} \ll \mu$ was needed to transition out of the SCBA regime into the FBA regime).

\section{Choosing parameters for numerics}\label{sec:numerics}

In order to perform numerical calculations, we need to carefully choose numerical values of $n_{\text{imp}}^{\pd}$ and $u_0^{\pd}$. To do so, we first obtain the monkey saddle and cusp singularities by tuning tight binding models, and then use the parameters of the models to choose $n_{\text{imp}}^{\pd}$ and $u_0^{\pd}$. (Recall from the discussion in Sec~\ref{sec:intro}, that we require $n_{\text{imp}}^{\pd} \ll n_{\text{el}}^{\pd}$ and $u_0^{\pd} \ll A_{\text{u.c}}^{\pd} \, E_s^{\,}$, where $A_{\text{u.c}}^{\pd}$ is the area of the real lattice unit cell). 

\subsection{Monkey saddle}

A single monkey saddle can be obtained in the Haldane model~\citep{Haldane_model} defined on the hexagonal lattice. The nearest neighbor vectors originating from the A sublattice on to the B sublattice are given by
\begin{equation}
\vb{a}_1^{\pd} = \begin{pmatrix} 1 \\ \\ 0 \end{pmatrix}, \vb{a}_2^{\pd} = \begin{pmatrix} -\frac{1}{2} \\ \\ \frac{\sqrt{3}}{2} \end{pmatrix} \& \, \vb{a}_3^{\pd} = \begin{pmatrix} -\frac{1}{2} \\ \\ -\frac{\sqrt{3}}{2} \end{pmatrix}.
\end{equation}

Here we have set the lengths to unity. The next nearest neighbor vectors are given by $\vb{b}_1^{\pd} = \vb{a}_2^{\pd} - \vb{a}_3^{\pd}$, $\vb{b}_2^{\pd} = \vb{a}_3^{\pd} - \vb{a}_1^{\pd}$ and $\vb{b}_3^{\pd} = \vb{a}_1^{\pd} - \vb{a}_2^{\pd}$. The full $\vb{k}$-space Hamiltonian is then the sum of nearest neighbor, staggered chemical potential and next nearest neighbor terms: $H(\vb{k}) = H_0^{\pd} (\vb{k}) + M \sigma_z^{\pd} + 2 t_2^{\pd} \sum_i \sin (\vb{k} \cdot \vb{b}_i^{\pd}) \, \sigma_z^{\pd}$, where $\sigma_z^{\pd}$ is the familiar Pauli matrix and
\begin{equation}
H_0^{\pd} (\vb{k}) = \begin{pmatrix}
0 & t_1^{\pd} \sum_i e^{i \vectorbold{k} \cdot \vectorbold{a}_i} \\
t_1^{\pd} \sum_i e^{-i \vectorbold{k} \cdot \vectorbold{a}_i} & 0
\end{pmatrix}.
\end{equation}
This Hamiltonian can be diagonalized exactly to give two bands indexed by $n = 1, 2$
\begin{multline}
\varepsilon_{n}^{\pd} (\vectorbold{k}) = (-1)^n \bigg[ \bigg(M + 2 t_2 \sum_i \sin (\vectorbold{k} \cdot \vectorbold{b}_i^{\pd}) \bigg)^2 \\ + t_1^2 \bigg| \sum_i e^{i \vectorbold{k} \cdot \vectorbold{a}_i^{\pd}} \bigg|^2 \bigg]^{\frac{1}{2}} .
\end{multline}
Consider the following, three-fold rotation consistent high symmetry point
\begin{equation}
\vb{K}_-^{\pd} = \frac{4\pi}{3\sqrt{3}} \begin{pmatrix}
0 \\ \\
1
\end{pmatrix}.
\end{equation}
We perform the following tuning of the staggered chemical potential
\begin{equation}
M \rightarrow \frac{t_1^2 - 18 \, t_2^2}{2 \, \sqrt{3} \, t_2^{\pd}}.
\end{equation}
Under this, the Taylor expanded upper band $(n=2)$ dispersion around the $K_-^{\pd}$ point reads
\begin{align}
\varepsilon_2^{\pd} (\vb{k} + \vb{K}_-^{\pd}) \approx & \frac{t_1^2}{2 \, \sqrt{3} \, |t_2^{\pd}|} + \frac{3 \, \sqrt{3} \, |t_2^{\pd}|}{2} \left( k_y^3 - 3 \, k_y^{\pd} \, k_x^2 \right) \nonumber \\
& + O \left( k^4 \right).
\end{align}
This is clearly the monkey saddle that we sought to obtain. Let us choose $t_1^{\pd} = 1$ and $t_2^{\pd} = 1/2$. By comparing the energy at $\vb{K}_-^{\pd}$ with the energy at the other high symmetry points viz. $\vb{K}_+^{\pd}$, $M$ and $\Gamma$ points, we can calculate $E_s^{\,}$, which we find to be approximately $1.68$. The area of the unit cell is $3\sqrt{3}/2$ so that we need
\begin{equation}
u_0^{\pd} \ll A_{\text{u.c}}^{\pd} \, E_s^{\pd} \approx 4.36.
\end{equation}
With two atoms per unit cell, the atomic density is $4/(3\sqrt{3}) \approx 0.77$ and we should choose $n_{\text{imp}}^{\pd} \ll 0.77$. For generating data for panels (a) and (b) in Fig~\ref{fig:DOS_Borns}, we used $n_{\text{imp}}^{\pd} = 0.1$ and $u_0^{\pd} = 1$. The overall coefficient $\alpha$ that multiplies the singularity was defined earlier in Appendix~\ref{app:scaling}. In the present case, it is found to be $\alpha = 3\sqrt{3} / 4$.

\subsection{Cusp}

One of the simplest models that yields the cusp singularity is a one band tight binding model defined on a square lattice, with asymmetric $x$ and $y$ nearest and next nearest neighbor hoppings $t_{1x}^{\pd}$, $t_{1y}^{\pd}$, $t_{2x}^{\pd}$ and $t_{2y}^{\pd}$. The dispersion takes the form
\begin{align}
\varepsilon (\vb{k}) =& -2 \, t_{1x}^{\pd} \cos k_x^{\pd} - 2 \, t_{1y}^{\pd} \cos k_y^{\pd} \nonumber \\
& - 2 \, t_{2x}^{\pd} \cos 2k_x^{\pd} - 2 t_{2y}^{\pd} \cos 2 k_y^{\pd}.
\end{align}
By setting $t_{2y}^{\pd} = t_{1y}^{\pd} / 4$ and $t_{2x}^{\pd} = 0$, and series expanding around the point $\vb{X} = (0, \pi)$, we obtain the cusp singularity
\begin{align}
\varepsilon (\vb{k} + \vb{X}) \approx \, & \frac{1}{8} \left( -15 + 6 \, k_x^2 - 2 \, t_{1y}^{\pd} \, \left( -6 + k_y^4 \right) \right) \nonumber \\ & + O(k^6).
\end{align}
Let us choose $t_{1y}^{\pd} = 3$, so that we get $\alpha = \beta = 3/4$. Computing the energy difference between the $\vb{X}$ point and the $M$ and $\Gamma$ points (i.e $(\pi, \pi)$ and $(0,0)$), we find $E_s^{\,} = 4$. The density of atoms is unity so that we need to choose $n_{\text{imp}}^{\pd} \ll 1$ and $u_0^{\pd} \ll 4$. For generating panels (b) and (d) of Fig~\ref{fig:DOS_Borns}, we set $n_{\text{imp}}^{\pd} = 0.1$ and $u_0^{\pd} = 0.75$.


\section{Evaluation of $I(\omega, \mu, z_0^{\,})$}
\label{app:deriv}
\begin{figure}[h]
 \centering
 \includegraphics[width=\columnwidth]{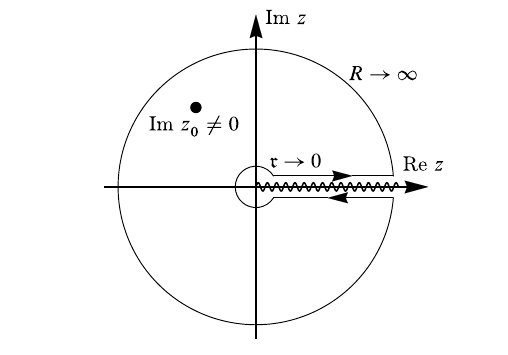}
 \caption{To evaluate the integral in Eq~\ref{eq:integral_common}, we use the contour depicted above. It has four distinct pieces: two circular contours respectively of radius $\mathfrak{r} \rightarrow 0$ and $R \rightarrow \infty$ and two contours slightly above and below the positive $x$ axis. It can be shown that the circular contours have zero contribution in their respective limits. The contours along the positive $x$ axis are separated by the branch cut and can be added and expressed in terms of the original integral of interest. We then apply residue theorem to evaluate this.}
 \label{fig:contour_complex}
\end{figure}

To evaluate the integral in Eq~\ref{eq:integral_common}, we split it into two parts, one for positive $\epsilon$ and one for negative $\epsilon$ and convert the negative integral by a substitution $z = - \epsilon$ to obtain
\begin{align}
I(\omega, \mu, z_0^{\pd}) = & -D_+^{\pd} \int_{0}^{\infty} dz \, z^{-\nu} \, \frac{1}{z - (\omega + \mu + z_0^{\pd})} \nonumber \\ & + D_-^{\pd} \int_{0}^{\infty} dz \, z^{-\nu} \, \frac{1}{z + (\omega + \mu + z_0^{\pd})}.
\end{align}
Since $\nu$ is a rational fraction, we need a branch cut to evaluate these integrals. We choose the $+x$ axis for this purpose and use the contour shown in Fig~\ref{fig:contour_complex}. Since the integrals respectively have poles at $z_0^{\pd} + \omega + \mu$ and $-(z_0^{\pd} + \omega + \mu)$, enclosed by the contours, we apply the residue theorem yielding ultimately
\begin{equation}
I(\omega, \mu, z_0^{\pd}) = \frac{2\pi i (\omega + \mu + z_0^{\pd})^{-\nu}}{1 - e^{-i 2\pi \nu}} \, [ e^{-i \pi \nu} D_-^{\pd} - D_+^{\pd} ].
\end{equation}
Notice that this is justified because $\text{Im} \, z_0^{\pd} \neq 0$ by our assumption and the poles will lie away from the $x$-axis, safely enclosed by the contour.

\section{Conductivity}
\label{app:finite}
To compute the conductivity using the Kubo formula for tight-binding models, we use the current operator defined as
\begin{equation}
\hat{\vbld{j}} =  \frac{e}{\mathcal{A}} \sum_{\vbld{k}} \nabla_{\vbld{k}}^{\pd} \xi_{\vbld{k}}^{\pd} \, \hat{c}_{\vbld{k}}^{\dag} \hat{c}_{\vbld{k}}.
\end{equation}
Although this lacks momentum or spatial dependence, we can use this current to obtain the essential features of the DC conductivity, including the conjoined effects of the singularity and impurities. To simplify the notation, we have made only the charge explicit in the above expression and have suppressed other dimensionfull factors (like mass). These can be reinstated later when necessary.

An essential component of this calculation is the retarded current correlation function that takes the form~\cite{bruus2004many,mahan}
\begin{align}
\Pi_{ij}^{R} (i q_n^{\pd}) = & \frac{e^2}{\mathcal{A}} \sum_{\vbld{p}} (\nabla_{\vbld{p}}^{\pd} \xi_{\vbld{p}}^{\pd})_{i}^{\pd} (\nabla_{\vbld{p}}^{\pd} \xi_{\vbld{p}}^{\pd})_{j}^{\pd} \nonumber \\
& \times \frac{1}{\beta} \sum_{m} \mathcal{G}(\vbld{p}, i p_m^{\pd} + iq_n^{\pd}) \, \mathcal{G} (\vbld{p}, i p_m^{\pd}),
\end{align}
where $\mathcal{A}$ is the area (or volume) and $i$ and $j$ are the directions indicating the component of the gradient to be used. The dc conductivity is then obtained by using the Lehmann representation for the Green's functions (in terms of the spectral function), performing the Matsubara sums, analytically continuing $i q_n^{\pd} \rightarrow \omega + i \delta$, and evaluating the $\omega \rightarrow 0$ limit to get~\cite{mahan}
\begin{align}
\sigma_{ij} =  e^2 \int \frac{d^2 p}{(2\pi)^2} \, \Bigg[ & \frac{\partial \xi_{\vbld{p}}^{\pd}}{\partial p_{i}^{\pd}} \, \frac{\partial \xi_{\vbld{p}}^{\pd}}{\partial p_{j}^{\pd}} \nonumber
\\ & \times \int \frac{d \epsilon}{2\pi} A^2_{\vbld{p}} (\epsilon) \left( - \frac{d n_{\text{F}}^{\pd}(\epsilon)}{d\epsilon} \right) \Bigg].
\end{align}
Let us assume that the inverse of the scattering lifetime $\tau_{\vb{p}}^{-1}(\omega, \mu)$ is finite and small. If we further assume that it is approximately independent of $\vb{p}$ and that the dependence on $\omega$ is ``weak" in comparison the strong $\omega^2$ of the spectral function $A_{\vbld{p}}^{\pd} ( \omega)$, we can altogether drop $\omega$ and $\vb{p}$ dependence so as to justify a relabelling as $\tau^{-1}(\mu)$ (since it still depends on the Fermi level). We briefly elaborate on the latter point, that is the weak $\omega$-dependence of $(\tau^{-1}(\omega,\mu))^2$ as compared to $\omega^2$, both of which appear in the denominator of the spectral function. We first note that within each of the Born approximations discussed above, $(\tau^{-1}(\omega,\mu))^2$ depends on $\omega$ and $\mu$ in the same way (more concretely, it depends on $\omega + \mu$). From panels (b), (d), (f) and (h) of Fig~\ref{fig:sigma_and_tauInv}, we see that as $\omega$ (or $\mu$ in the figure) changes from $0.1$ to $1$ in the numerical calculations, $\tau_{\vb{p}}^{-1}(\omega, \mu)$ at the most doubles in value. Thus, a change of $\omega^2$ by two orders of magnitude corresponds to a fourfold change in $(\tau^{-1}(\omega,\mu))^2$, which still has a value much smaller than unity. Using this property, we drop the $\omega$ dependence of $\tau^{-1}(\omega,\mu)$. The spectral function then effectively becomes a Cauchy distribution. We can then use the properties of Cauchy distribution and the delta function to approximate the above expression as~\cite{mahan}
\begin{equation}
\sigma_{ij}^{\pd} = 2 e^2 \, \tau(\mu) \int \frac{d^2 p}{(2\pi)^2} \, \frac{\partial \xi_{\vbld{p}}^{\pd}}{\partial p_{i}^{\pd}} \, \frac{\partial \xi_{\vbld{p}}^{\pd}}{\partial p_{j}^{\pd}} \, \delta (\xi_{\vb{p}}^{\pd} - \mu).
\label{eq:conductivity}
\end{equation}
At this point we have just recovered the conventional expression for conductivity that is proportional to the electron lifetime at the Fermi level. Nevertheless, the main outstanding issue is to ensure that the integral is finite in the $\mu \rightarrow 0$ limit for HOVHS, since we are dealing with systems that have infrared divergences in the clean limit. We do this below, where we show that the integral is finite and non-zero in the $\mu \rightarrow 0$ limit for the cusp and monkey saddle singularities.

\subsection{Cusp}
\begin{figure}[h]
 \centering
 \includegraphics[width=\columnwidth]{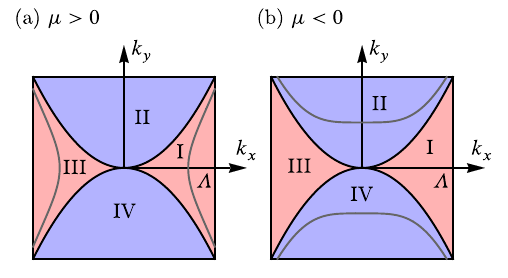}
 \caption{The constant energy contours of the cusp singularity, for a positive and negative energy are respectively depicted in (a) and (b). The region with energy below the singularity is shaded blue, while the region with energy above the singularity is shaded red. The cutoff procedure we will use is a box cutoff with $\Lambda$ for $k_x^{\,}$ and $\Lambda^2$ for $k_y^{\,}$.}
 \label{fig:cusp_cutoff}
\end{figure}

The cusp singularity has a canonical dispersion $\xi_{\vbld{k}}^{\pd} = k_x^4 - k_y^2$. To evaluate the integral in Eq~\ref{eq:conductivity} we will use a box boundary with a $k_x^{\pd}$- cutoff $\Lambda$ and a corresponding $k_y^{\pd}$ cutoff $\Lambda^2$ as shown in Fig~\ref{fig:cusp_cutoff}. First let us consider the case $\mu > 0$. Here we can restrict the integration to the the regions $\RN{1}$ and $\RN{3}$. The integral then takes the form
\begin{equation}
I_{ij}^{\pd}(\mu, \Lambda) = \int\limits_{-\Lambda}^{\Lambda} \frac{d k_x^{\pd}}{(2\pi)} \int\limits_{-k_x^2}^{k_x^2} \frac{d k_y^{\pd}}{(2\pi)}  (\partial_{i}^{\pd} \xi ) (\partial_{j}^{\pd} \xi ) \, \delta \left( k_x^4 - k_y^2 - \mu \right). 
\end{equation}
The delta function can be expressed as
\begin{align}
\delta \bigg( k_x^4 & - k_y^2 - \mu \bigg) = \Theta \left( k_x^4 - \mu \right) \nonumber \\
& \times \left[ \frac{\delta \left( k_y^{\pd} - \sqrt{k_x^4 - \mu} \right) }{|-2 k_y^{\pd}|} + \frac{\delta \left( k_y^{\pd} + \sqrt{k_x^4 - \mu} \right) }{|-2 k_y^{\pd}|} \right] \label{Eq:delta_idty_cusp}
\end{align}

The step function ensures that the roots are real (they always lie in the integration domain when real). Now consider $i = x$ and $j = x$. We can easily evaluate the $k_y^{\pd}$ integral using the above delta function identity to obtain
\begin{subequations}
\begin{align}
I_{x x}^{\pd} (\mu, \Lambda) = & \int\limits_{-\Lambda}^{\Lambda} \frac{d k_x^{\pd}}{(2\pi)^2} \frac{16 \, k_x^6}{\sqrt{k_x^4 - \mu}} \, \Theta \left( k_x^4 - \mu \right) \\
= & \frac{8}{\pi^2} \int\limits_{\mu^{1/4}}^{\Lambda} d k_x^{\pd} \frac{k_x^6}{\sqrt{k_x^4 - \mu}}.
\end{align}\label{Eq:cusp_intg_x}
\end{subequations}
At this point it is clear why we needed the UV cutoff $\Lambda$. Without it, the integral in Eq~\ref{Eq:cusp_intg_x}b diverges. We now make the substitution $t = \mu/ k_x^4$. The integral becomes
\begin{equation}
I_{x x}^{\pd} = -\frac{2 \mu^{5/4}}{\pi^2} \int\limits_{1}^{\mu / \Lambda^4} dt \, \, t^{-9/4} (1 - t)^{-1/2} .
\end{equation}
This integral can be easily rewritten in the form of an incomplete Beta function, using the substitution $l = 1 - t$:
\begin{subequations}
\begin{align}
I_{x x}^{\pd} =& \frac{2 \mu^{5/4}}{\pi^2} \int\limits_{0}^{1 - \mu/\Lambda^4} dl \,\, l^{1/2 - 1} (1 - l)^{-5/4 - 1} \\
=& \frac{2 \mu^{5/4}}{\pi^2} B_{1-\mu/\Lambda^4}^{\pd} (1/2, -5/4).
\end{align}
\end{subequations}
We now series expand this around $\mu = 0$:
\begin{align}
I_{xx}^{\pd} (\mu, \Lambda) \approx  \, \Lambda^5 \bigg( & 0.16 + 0.41 \, \mu / \Lambda^4 -0.29 \, (\mu / \Lambda^4 )^{5/4}  \nonumber \\  
& + O \left( (\mu / \Lambda^4 )^{2} \right) \bigg).
\end{align}
Thus, in the $\mu \rightarrow 0+$ limit, the integral goes to a finite constant with a small linear correction to leading order.  For $i = x$ and $j = y$ (or vice versa), the sum over the two roots $k_y^{\pd} = -\sqrt{k_x^4 - \mu}$ and $k_y^{\pd} = +\sqrt{k_x^4 - \mu}$, causes the $k_y$ integral to vanish since $(\partial_x^{\pd} \xi)(\partial_y^{\pd} \xi) = -8 k_x^3 k_y^{\pd}$ is odd in $k_y^{\pd}$. Therefore we continue to investigate the case $i = y$, $j = y$. After applying the identity in Eq~\ref{Eq:delta_idty_cusp} and simplifying, the integral in this case reads
\begin{subequations}
\begin{align}
I_{yy}^{\pd} (\mu, \Lambda) =& \frac{1}{\pi^2} \int\limits_{-\Lambda}^{\Lambda} d k_x^{\pd} \, \sqrt{k_x^4 - \mu} \\
=& \frac{2}{\pi^2} \int\limits_{\mu^{1/4}}^{\Lambda} d k_x^{\pd} \, \sqrt{k_x^4 - \mu}
\end{align}
\end{subequations}
Once again we use the substitution $x^4 = \mu / t$ to obtain
\begin{equation}
I_{yy}^{\pd} (\mu, \Lambda) = - \frac{\mu^{3/4}}{2 \pi^2} \int\limits_{1}^{\mu/\Lambda^4} dt \, \, t^{-7/4} (1-t)^{1/2}.
\end{equation}
The substitution $t = 1 - l$ puts this in the form of an incomplete Beta function
\begin{subequations}
\begin{align}
I_{yy}^{\pd} (\mu, \Lambda) = & \frac{\mu^{3/4}}{2 \pi^2} \int\limits_{0}^{1 - \mu/\Lambda^4} dl \,\, l^{3/2 - 1} (1 - l)^{-3/4 - 1} \\
= & \frac{\mu^{3/4}}{2 \pi^2} B_{1 - \mu/\Lambda^4}^{\pd} (3/2, -3/4)
\end{align}
\end{subequations}
We can now expand around $\mu = 0$ 
\begin{subequations}
\begin{align}
I_{yy} (\mu, \Lambda) \approx \,\, \Lambda^3 \bigg( & 0.07 - 0.18 \, (\mu / \Lambda^4)^{3/4}  + 0.10 \, \mu / \Lambda^4  \nonumber \\ 
&+ O \left( \left( \mu / \Lambda \right)^{7/4} \right) \bigg)
\end{align}
\end{subequations}

We examine $\mu < 0$. The contours for this case are shown in Fig~\ref{fig:cusp_cutoff}(b). The cutoff is implemented through the $k_y^{\pd}$ integral that has limits $\pm \Lambda^2$. The delta function identity to use in this case is
\begin{align}
\delta \bigg( k_x^4 & - k_y^2 - \mu \bigg) = \Theta \left( \Lambda^2 - \sqrt{x^4 + |\mu|} \right) \nonumber \\
& \times \left[ \frac{\delta \left( k_y^{\pd} - \sqrt{k_x^4 + |\mu|} \right) }{|-2 k_y^{\pd}|} + \frac{\delta \left( k_y^{\pd} + \sqrt{k_x^4 + |\mu|} \right) }{|-2 k_y^{\pd}|} \right] \label{Eq:delta_idty_cusp}
\end{align}
Although the roots are always real (since $\mu < 0$), we need the step function to ensure that they fall within the integral domain, i.e $\tilde{k}_{y}^{\pm} = \pm \sqrt{k_x^4 + |\mu|} \in (-\Lambda^2, \Lambda^2)$. This leads to a corresponding restriction on $k_x^{\pd}$ as $|k_x^{\pd}| < (\Lambda^4 - |\mu|)^{1/4}$. Let us treat $I_{xx}^{\pd}$ first. After some simplification, we get
\begin{equation}
I_{xx}^{\pd} (\mu, \Lambda) = \frac{8}{\pi^2} \int\limits_{0}^{(\Lambda^4 - |\mu|)^{1/4}} d k_x^{\pd} \frac{k_x^6}{\sqrt{k_x^4 + |\mu|}}.
\end{equation}
By a series of substitutions, first $t = |\mu| / (k_x^4 + |\mu|)$ and then $l = 1- t$ we get
\begin{subequations}
\begin{align}
I_{xx}^{\pd} (\mu, \Lambda) = & \frac{2 |\mu|^{5/4}}{\pi^2} \int\limits_{0}^{1-|\mu|/\Lambda^4} dl \,\, l^{7/4-1} (1-l)^{-5/4 - 1} \\
=& \frac{2 |\mu|^{5/4}}{\pi^2} B_{1-|\mu|/\Lambda^4} (7/4, -5/4).
\end{align}
\end{subequations}
Series expanding this, we obtain to leading order in the $\mu \rightarrow 0$ limit
\begin{align}
I_{xx}^{\pd} (\mu, \Lambda) \approx \Lambda^5 \bigg( & 0.16 - 0.61 \, |\mu| / \Lambda^4 + 0.41 \, \left( |\mu| / \Lambda^4 \right)^{5/4} \nonumber \\
& + O \left( \left( |\mu| / \Lambda^4 \right)^{2} \right) \bigg).
\end{align}
Lastly, we compute $I_{yy}^{\pd}$. Here again we first make the substitution $t = |\mu|/(k_x^4 + |\mu|)$ followed by the substitution $l = 1 - t$ to get
\begin{subequations}
\begin{align}
I_{yy}^{\pd} (\mu, \Lambda) =& \frac{ |\mu|^{3/4}}{2 \pi^2} \int\limits_{0}^{1 - |\mu|/\Lambda^4} dl \,\, l^{1/4 - 1} (1 - l)^{1/4 - 1} \\
=& \frac{ |\mu|^{3/4}}{2 \pi^2} B_{1 - |\mu|/\Lambda^4} (-1/4, -1/4).
\end{align}
\end{subequations}
Expanding this to leading order in $|\mu|$ we get 
\begin{align}
I_{yy} (\mu, \Lambda) \approx \,\, \Lambda^3 \bigg( & 0.07 + 0.25 \, (\mu / \Lambda^4)^{3/4}  - 0.15 \, \mu / \Lambda^4  \nonumber \\ 
&+ O \left( \left( \mu / \Lambda \right)^{7/4} \right) \bigg)
\end{align}
We see that integral is not symmetric between $\mu < 0$ and $\mu > 0$ although both cases have the same $\mu \rightarrow 0$ limit. Also $I_{xx}^{\pd} (\mu, \Lambda)$ and $I_{yy}^{\pd} (\mu, \Lambda)$ are not equal. In fact we have in the $\mu \rightarrow 0$ limit
\begin{equation}
\frac{I_{xx}^{\pd} (\mu, \Lambda)}{I_{yy}^{\pd} (\mu, \Lambda)} \, \xrightarrow{\mu \rightarrow 0} \, \frac{12 \Lambda^2}{5}.
\end{equation}

\subsection{Monkey saddle}
\begin{figure}[h]
 \centering
 \includegraphics[width=\columnwidth]{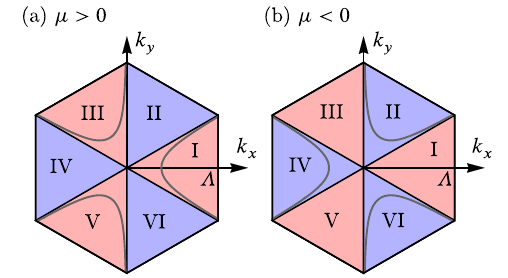}
 \caption{In evaluating the integral in Eq~\ref{eq:integral_common}, the monkey saddle requires a careful choice of UV cutoff since the box cutoff makes analytic treatment difficult. As before, we color the regions with energy below and above the singularity respectively blue and red. For any energy $\mu$, there are three disconnected contours that are related by $2\pi / 3$ rotations. This serves as the motivation for using the hexagonal cutoff scheme for the integral.}
 \label{fig:monkey_cutoff}
\end{figure}

We begin by noting the following
\begin{align}
& \xi_{\vbld{k}}^{\pd} =: \xi (k_x^{\pd}, k_y^{\pd}) = k_x^3 - 3 k_x^{\pd} k_y^2, \\
\implies \, & \frac{\partial \xi_{\vbld{k}}^{\pd}}{\partial k_x^{\pd}} = 3 k_x^2 - 3 k_y^2, \\
& \frac{\partial \xi_{\vbld{k}}^{\pd}}{\partial k_y^{\pd}} = -6 k_x^{\pd} k_y^{\pd} .
\end{align}
The constant energy contours at an energy $\mu$ are 
\begin{equation}
k_y^{\pd} = \pm \sqrt{\frac{k_x^3 - \mu}{3 k_x^{\pd}}}.
\end{equation}
For $\mu > 0$, there are three disconnected pieces that are related by a $2\pi / 3$ rotation. These reside respectively in regions $\RN{1}$, $\RN{3}$ and $\RN{5}$ in Fig~\ref{fig:cusp_cutoff}(a). Thus for $\mu > 0$, the integral can be restricted to these regions. In region $\RN{1}$, a cutoff $\Lambda$ on $k_x^{\pd}$ serves as a cutoff on $k_y$ as well, so that the region of integration becomes bounded (a triangular region to be precise). However in $\RN{3}$ and $\RN{5}$, a cutoff on $k_x^{\pd}$ does not automatically restrict the $k_y^{\pd}$ integral. This is because the $k_x^{\pd} \rightarrow 0$ limit corresponds to $k_y^{\pd} \rightarrow \pm \infty$. Now we might attempt to additionally put a hard cutoff on $k_y^{\pd}$, say $\Lambda'$. But this will also make the range of $k_x^{\pd}$ integration restricted and dependant on $\mu$ and $\Lambda'$ (since every $k_x^{\pd}$ will not give a valid $k_y^{\pd}$ on the $\mu$-contour). More precisely, the lower and upper limits of the $k_x^{\pd}$ integral will be the solutions of
\begin{equation}
k_x^3 - 3 k_x \Lambda'^2 = \mu.
\end{equation} 
This is obviously not easy to work with. We therefore use a different cutoff procedure, with a hexagonal region as depicted in Fig~\ref{fig:monkey_cutoff}. The reason for this is that, regions $\RN{3}$ and $\RN{5}$ can be `transformed' into region $\RN{1}$ by simple rotations of respectively $-2\pi/3$ and $2\pi/3$. Such rotations will not change the dispersion, but only the derivatives $\partial_{i}^{\pd} \xi$ and $\partial_{j}^{\pd} \xi$, that change by linear combinations of $\partial_{x}^{\pd} \xi$ and $\partial_{y}^{\pd} \xi$ (By the chain rule since we  have a linear transformation of the coordinates). 

Let us call the rotated variables $u$ and $v$. Let us $\rho_{\theta}^{\pd}$ denote the rotation matrix for angle $\theta$. As mentioned above, we have $\xi (u, v) = \xi (\rho_{\theta}^{\pd} (k_x^{\pd}, k_y^{\pd})) = \xi (k_x^{\pd}, k_y^{\pd})$ but the derivatives become
\begin{align}
\frac{\partial \xi}{\partial k_x^{\pd}} & \xrightarrow{\rho_{\pm 2\pi/3}^{\pd}} -\frac{3}{2} (u^2 - v^2 \pm 2 \sqrt{3} \, uv) \\
\frac{\partial \xi}{\partial k_y^{\pd}} & \xrightarrow{\rho_{\pm 2\pi/3}^{\pd}} \mp \frac{3 \sqrt{3}}{2} (u^2 - v^2 \mp 2\, uv/ \sqrt{3} )
\end{align}
Now the integrals over the three regions can be combined into a single integral over region $\RN{1}$ using this scheme. We now compute $(\partial_{i}^{\pd} \xi) (\partial_{j}^{\pd} \xi)$ summed over the three regions
\begin{align}
& (\partial_{x}^{\pd} \xi)^2|_{\RN{1}}^{\pd} + (\partial_{x}^{\pd} \xi)^2|_{\RN{2}}^{\pd} + (\partial_{x}^{\pd} \xi)^2|_{\RN{3}}^{\pd}  \rightarrow \frac{27}{2} (u^2 + v^2)^2 \\
& (\partial_{x}^{\pd} \xi)(\partial_{y}^{\pd} \xi)|_{\RN{1}}^{\pd} + (\partial_{x}^{\pd} \xi)(\partial_{y}^{\pd} \xi)|_{\RN{2}}^{\pd} + (\partial_{x}^{\pd} \xi)(\partial_{y}^{\pd} \xi)|_{\RN{3}}^{\pd} \rightarrow 0 \\
& (\partial_{y}^{\pd} \xi)^2|_{\RN{1}}^{\pd} + (\partial_{y}^{\pd} \xi)^2|_{\RN{2}}^{\pd} + (\partial_{y}^{\pd} \xi)^2|_{\RN{3}}^{\pd}  \rightarrow \frac{27}{2} (u^2 + v^2)^2
\end{align}
We have to thus evaluate the following integral
\begin{equation}
I(\mu, \Lambda) = \frac{27}{2} \int\limits_{0}^{\Lambda} \frac{du}{(2\pi)} \int\limits_{-\sqrt{3} x}^{\sqrt{3} x} \frac{dv}{(2\pi)} (u^2 + v^2)^2 \delta (u^3 - 3 u v^2 - \mu)
\end{equation}
We proceed as in the previous section, first by using a delta function identity to calculate the $v$ integral then use appropriate substitution to recast it in terms of the incomplete Beta functions. We obtain
\begin{align}
I(\mu, \Lambda) =& \frac{2}{\sqrt{3} \pi^2} \, \frac{\mu}{\Lambda^3} \, B_{1 - \mu/\Lambda^3}^{\pd} \left( \frac{1}{2}, -1 \right) \nonumber \\
& - \frac{1}{\sqrt{3} \pi^2} \, \frac{\mu}{\Lambda^3} \, B_{1- \mu / \Lambda^3} \left( \frac{1}{2}, 0 \right) \nonumber \\
& + \frac{1}{4 \sqrt{3} \pi^2} \, \frac{\mu}{\Lambda^3} \sqrt{1- \frac{\mu}{\Lambda^3}}.
\end{align}
The series expansion around $\mu = 0$ reads
\begin{align}
I(\mu, \Lambda) \approx & \, 0.12 - 0.04 \, (\mu / \Lambda^3) - 0.02 \, (\mu / \Lambda^3)^2 \nonumber \\
& + O(\, (\mu / \Lambda^3)^3).
\end{align}
It is easy to see by a simple change of variables that this derivation and final result hold for $\mu < 0$ as well.

\bibliographystyle{apsrev4-1}

%
\end{document}